\documentclass[aps,prx,twocolumn,floatfix,longbibliography,nofootinbib]{revtex4-2}
\usepackage{tikz}
\usepackage[utf8]{inputenc}
\usepackage{natbib}
\usepackage{graphicx}
\usepackage{xcolor}
\usepackage[tbtags]{amsmath}
\usepackage[colorlinks, linkcolor=blue]{hyperref}
\hypersetup{colorlinks,allcolors=black}
\usepackage{amssymb}
\usepackage{gensymb}
\usepackage{txfonts}
\usepackage{comment}
\usepackage{float}
\usepackage{amsmath}
\usepackage{tabularx,graphicx}
\usepackage{epstopdf}
\usepackage{latexsym}
\usepackage{color, colortbl}
\usepackage{psfrag}
\usepackage{bbm}
\usepackage{bm}
\usepackage{titlesec}
\usepackage{dsfont}
\usepackage{feynmp}
\usepackage{slashed}
\usepackage{multirow}
\usepackage[normalem]{ulem}
\renewcommand{\vec}[1]{\mathbf{#1}}

\usepackage{bm}  % in the preamble
\renewcommand{\vr}{\bm{r}}
\newcommand{\vk}{\bm{k}}

\newcommand{\vQ}{\bm{Q}}
\newcommand{\ve}{\bm{e}}

\def \beq {\begin{eqnarray}}
\def \eeq {\end{eqnarray}}
\def \tn {\textnormal}

\newcommand{\triucell}{\tikz[baseline=-0.5ex,scale=0.2]{
  \draw (0-0.5,0-0.5) -- (1-0.5,0-0.5) -- (3/2-0.5,1.73/2-0.5) -- (1/2-0.5,1.73/2-0.5) -- (0-0.5,0-0.5) -- (1-0.5,0-0.5) -- (1/2-0.5,1.73/2-0.5) -- cycle;
}}

\renewcommand{\appendix}{%
  \par
  \setcounter{section}{0}%
  \renewcommand{\thesection}{\Alph{section}}% Alphabetic section numbering
  \renewcommand{\theHsection}{appendix.\Alph{section}}% Unique hyperlink anchor
  \titleformat{\section}[block]{\normalfont\bfseries\centering}{Appendix \thesection:}{1em}{}%
}
\makeatother
\begin{document}
\title{\bf SU(4) Kondo Lattice in Semiconductor Moir\'e Materials}

\author{Sunghoon Kim}\email{ sk3299@cornell.edu}
\affiliation{Department of Physics, Cornell University, Ithaca, New York 14853, USA.}
\begin{abstract}
Motivated by recent advances in transition metal dichalcogenide (TMD) moir\'e materials, we propose TMD moir\'e multilayers as a platform for realizing an approximately SU(4)-symmetric triangular Kondo lattice, generalizing the concept of the double quantum dot model. Our model extends the conventional Kondo lattice by incorporating a three-site exchange of SU(4) local moments, which drives spontaneous time-reversal and lattice symmetry breaking. Using a parton mean-field approach, we map out the phase diagram as a function of three-site exchange and hole doping. In the Kondo-unscreened regime, we identify Mott insulating phases, including bond-ordered states and a chiral spin liquid. With increasing doping, Kondo hybridization gives rise to a heavy Fermi liquid that exhibits distinct patterns of lattice symmetry breaking, with or without topological responses. We conclude with directions for future study.
\end{abstract}

\maketitle
\textit{Introduction.}-
Transition metal dichalcogenide (TMD) moiré materials have emerged as an ideal platform for studying a variety of strongly correlated phenomena \cite{Wu_Hubbard_PRL,Wu_QSH_PRL,Mak_Shan_review}. TMD bilayers with layer asymmetry can simulate a single-band Hubbard model on a triangular moiré superlattice, with bandwidth tunable by an external electric field. The strongly interacting regime is experimentally accessible, as demonstrated by the emergence of correlated insulators at commensurate fillings \cite{Tang_simulation_2020,Regan_mottwigner_2020,Xu_mottwigner_2020,Li_imagewigner_2021}, as well as bandwidth-tuned continuous metal-insulator transitions at half filling \cite{li_continuous_2021,ghiotto_quantum_2021}. In this regime, exotic quantum phases such as quantum spin liquids \cite{Savary_2016,QSL_RMP,Broholm_2020} may be realized, though probing them remains challenging since their low-energy excitations do not directly couple to external probes.

Beyond these correlated insulators, coupling localized moments to itinerant carriers via exchange interactions can give rise to Kondo lattices, opening another avenue for rich many-body physics \cite{Doniach_1977,HFL_RMP,Coleman_Si_HFLreview,Senthil_FL*_PRL,Senthil_FL*_PRB}. A striking experimental advance in this direction is the realization of gate-tunable Kondo lattices in TMD moir\'e bilayers \cite{Zhao_Kondo23}. In MoTe$_2$/WSe$_2$ heterobilayers, the moir\'e potential is stronger in the MoTe$_2$  layer, resulting in a flatter valence band than that of the WSe$_2$ layer. When the MoTe$_2$ layer is half-filled, local moments emerge at a total commensurate filling $\nu_T=1$. Therefore, a Kondo lattice can be realized when the MoTe$_2$ layer remains half-filled while the WSe$_2$ layer is gradually doped. This platform has revealed many interesting phenomena, including a density-tuned Kondo breakdown \cite{Zhao_2024}, a Chern metal near the Kondo breakdown \cite{zhao_chernmetal}, and topological mixed-valence (and Kondo) insulators \cite{Han_TKI}.
Considerable theoretical work has also been devoted to investigating Kondo physics in TMD moiré materials \cite{Kuma_PRB_Kondo,Ruhman_Kondo,Guerci_2023,Qimiao_Kondo,Kim_TMVI,Guerci_TKI,Qimiao_TKS}.

Given the manifestations of rich Kondo physics in TMD moir\'e materials, it is natural to explore related platforms that may host even richer phenomena. One promising direction is the realization of Kondo lattices with an enlarged number of degrees of freedom, that is, Kondo lattices exhibiting SU($N>2$) symmetry \cite{Coleman_largeN,Read_Doniach_largeN,Auerbach_largeN,Millis_largeN,Assaad_SUN_Kondo}. Such models are particularly appealing since SU($N>2$) spins feature enhanced spin fluctuations that can stabilize exotic ground states \cite{Affleck_largeN_1,Affleck_largeN_2,Read_Sachdev_1989,Read_Sachdev_PRL1989,Read_Sachdev_PRB1990,Read_Sachdev_PRL1991,Sachdev_PRB1992,SCZhang_SO5,Assaad_SUN_QMC,Arovas_SUNsinglet,Xu_SU4Plaquette,Hermele_SU4plaquette,Gorshkov_Cold_SUN,Assaad_SUN_plaquette,Keselman_su4_prl,SU3_pairedCSL}. In graphene-based moir\'e materials, SU(4) and SU(8) Kondo lattice models have been proposed in twisted trilayer \cite{Lado_SU4TTG} and bilayer graphene \cite{SDS_SU8TBG}, with the SU(8) case in magic-angle twisted bilayer graphene rooted in a topological heavy fermion picture \cite{Song_Bernevig_THF}. In TMD moir\'e materials, a natural strategy is to exploit layer degrees of freedom \cite{zhang2020electrical,zhang_prl,Wilhelm_PRX_PSL,Boran_Yahui_SU4,Lorenzo_PRR}. Recently, Coulomb-coupled TMD bilayers with suppressed interlayer tunneling was theoretically proposed as a platform for simulating an SU(4) Hubbard model \cite{zhang_prl}, followed by experimental evidence in twisted AB-stacked homobilayers \cite{SU4_tWSe2_Expt}. 

%Increasing the number of flavors can enhance the stability of exotic ground states such as chiral spin liquids, and also raises the Kondo temperature. 

\begin{figure}[pth!]
\centering
\includegraphics[width=0.99\linewidth]{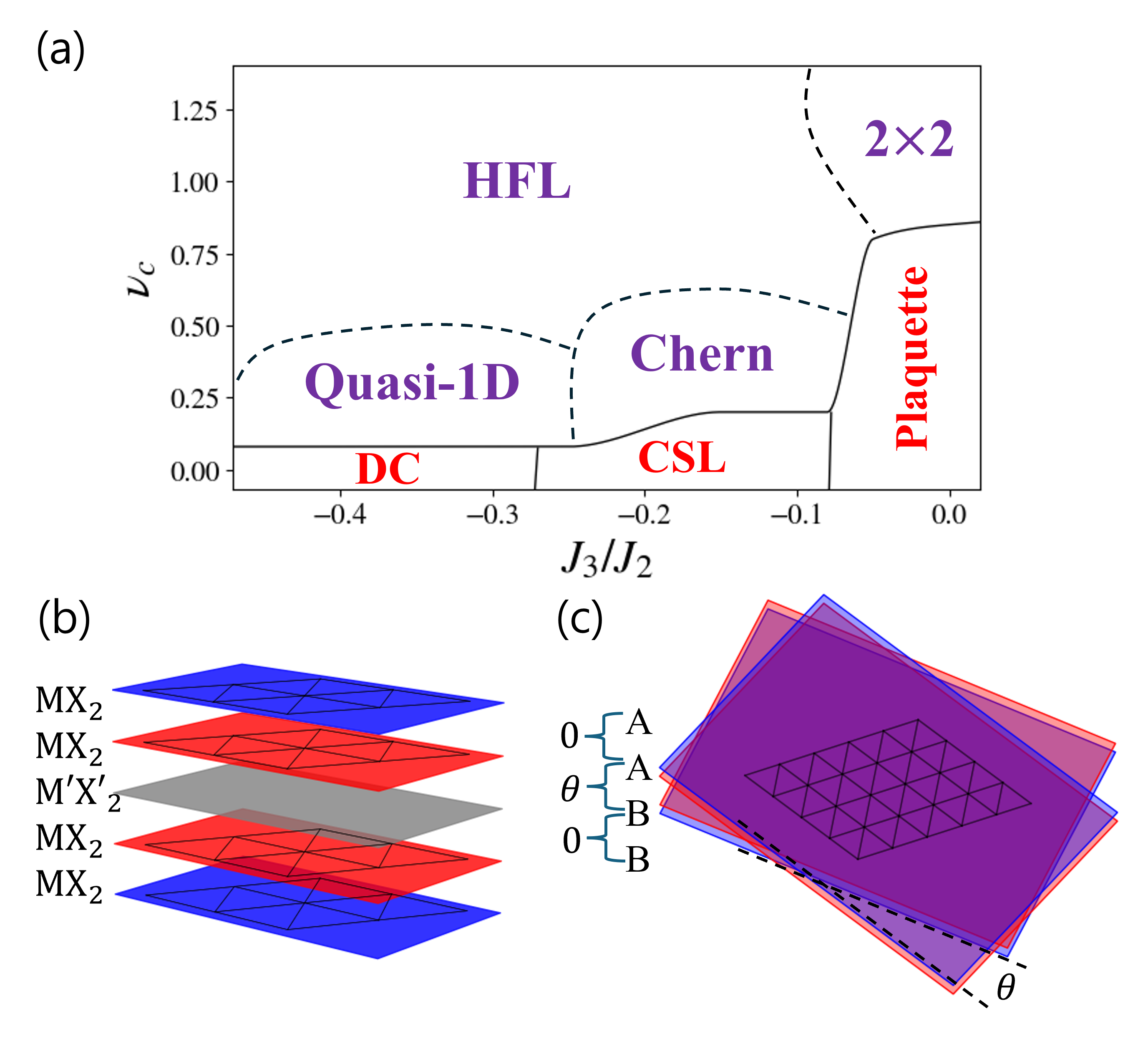} %0.9
\caption{\textbf{Schematics of SU(4) Kondo lattices and phase diagram.} (a) Schematic phase diagram at $\nu_f=1$ as a function of three-site exchange $J_3$ and $c$-hole filling $\nu_c$, with two-site exchange $J_2$ set to unity. Below the critical filling, the $f$-fermions are decoupled from the $c$-holes and form Mott insulators that host various bond-ordered phases---plaquette order, decoupled chain (DC)---as well as a chiral spin liquid (CSL). Above the critical filling, a Kondo transition to a heavy Fermi liquid occurs (red $\rightarrow$ purple). The heavy Fermi liquid (HFL) typically features various lattice symmetry breakings, with or without an anomalous Hall response. (b) TMD pentalayer Kondo lattice consisting of four identical layers and one distinct layer. The two layers adjacent to the middle layer are more correlated (red, $f$-layers) than the two outer layers (blue, $c$-layers). (c) Twisted AABB-stacked TMD tetralayer. A small twist angle $\theta$ is applied between the second (A) and third (B) layers. As in (b), the middle layers are more correlated than the outer layers.}
\label{fig:model}
\end{figure}

In this work, we extend these studies by proposing TMD moir\'e multilayers as a platform for realizing an SU(4)-symmetric triangular Kondo lattice. Our model goes beyond the conventional Kondo lattice model by incorporating a three-site exchange of SU(4) local moments, which induces spontaneous breaking of time-reversal and lattice symmetries. Using a parton mean-field approach, we map out the quantum phase diagram as a function of three-site exchange and doping, as illustrated schematically in Fig.~\ref{fig:model}(a). Below critical doping, we identify various Mott insulators including a chiral spin liquid and crystallized bond-ordered phases. Above critical doping, a heavy Fermi liquid with broken lattice symmetry emerges, with or without anomalous Hall responses. We expect our mean-field results to be reasonably reliable, since fluctuations around the saddle point are relatively suppressed in the SU(4) case. Our theoretical
study highlights that the TMD moir\'e SU(4) Kondo lattice can serve as a rich playground for realizing a plethora of correlated and topological phases.

\textit{SU(4) Kondo lattice in TMD multilayers.}-
Our model generalizes the concept of double quantum dot systems to TMD multilayers. A double quantum dot system consists of two adjacent quantum dots, each attached to separate metallic leads. When interdot coupling is strong while interdot tunneling remains weak, an SU(4)-symmetric system can be realized \cite{Borda_DoubleQD}. We extend this idea by replacing each dot by a strongly-correlated TMD ($f$-)layer, and each lead by a weakly-correlated TMD ($c$-)layer. There are two prominent approaches to construct a bilayer consisting of an $f$-layer and a $c$-layer: (i) using heterobilayers, such as MoTe$_2$/WSe$_2$ \cite{Zhao_Kondo23}, or (ii) placing the source of the moir\'e potential on one side \cite{Kuma_PRB_Kondo}. In this work, we focus on the second approach, and will comment on the first approach in the Outlook.

Building on the idea of the double quantum dot model, as well as established theoretical settings of the SU(4) Hubbard model \cite{zhang_prl} and TMD-based moir\'e Kondo lattices \cite{Kuma_PRB_Kondo}, we propose that an SU(4) Kondo lattice can be realized in the following two settings. The first setting consists of four identical TMD layers (MX$_2$) and one distinct TMD layer (M$'$X$'$$_2$) sandwiched by the second and the third MX$_2$ layers, as shown in Fig.~\ref{fig:model}(b). The sandwiched layer plays two distinct roles. First, it provides triangular moir\'e lattices (due to lattice mismatch) with different moir\'e potentials, thereby producing two sets of $f$- and $c$-layers. Second, it serves as an insulating barrier that suppresses the interlayer tunneling between the two $f$-layers. The four identical layers are atomically aligned so that the triangular lattices are aligned. In the strongly interacting regime, the $f$-layers form a Mott insulating phase at commensurate filling of one hole per site ($\nu_f=1$). Upon doping the $c$-layers, a Kondo lattice can be realized. We note that twisted hexagonal boron nitride (hBN) layers may also be used as the middle layer to produce the moir\'e potential \cite{Kim_thBN_2023}. 

The second setting is twisted AABB-stacked homobilayers, motivated by experimentally established twisted AB-stacked homobilayers \cite{SU4_tWSe2_Expt}. AB-stacking combined with strong spin–orbit coupling ensures that interlayer tunneling is spin-forbidden. For a small additional twist angle $\theta$, triangular moir\'e lattices emerge. Our setting extends this structure by adding an extra A- and B-layer on the top and bottom, as shown in Fig.~\ref{fig:model}(c). The outer $c$-layers are less correlated than the two inner $f$-layers.

\begin{figure*}[pth!]
\centering
\includegraphics[width=\linewidth]{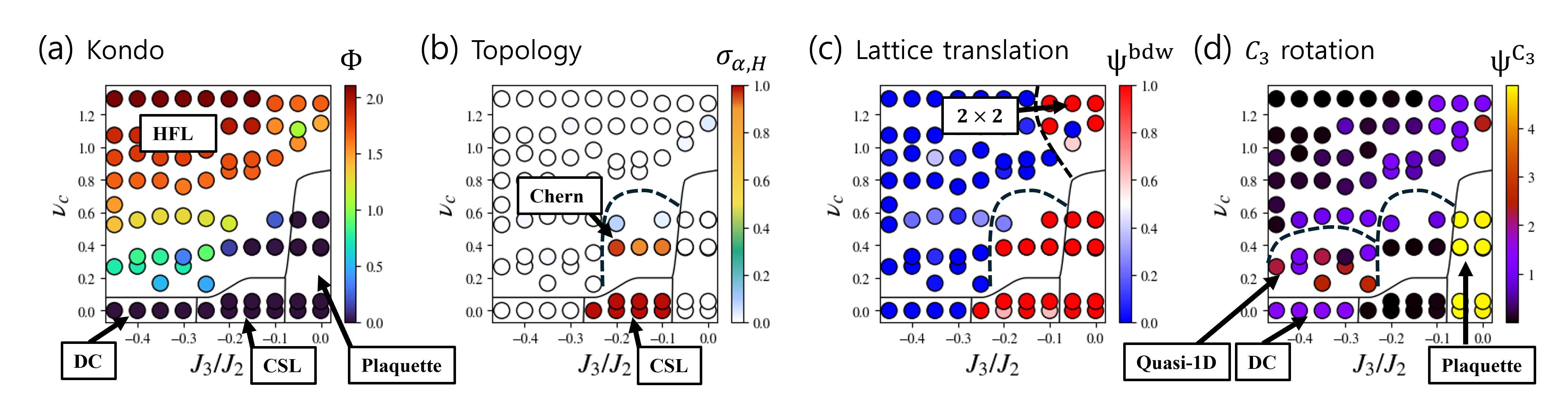} %0.9
\caption{\textbf{Self-consistent mean-field phase diagram at $\nu_f=1$ as a function of $J_3/J_2$ and $\nu_c$}. The phase diagram is obtained for a $12\times12$ system with $(t_c/J_2,J_K/J_2)=(-3,2)$. (a) Density-tuned Kondo transition. Color coding represents the Kondo hybridization order parameter $\Phi$. In the Kondo-unscreened regime, the $f$-fermions form various Mott insulating phases: DC, CSL, and plaquette order.
(b) Topological phase transition. Color coding represents the flavor Hall conductivity $\sigma_{\alpha,H}$. In the Kondo-unscreened regime, $\sigma_{\alpha,H}$ is quantized to unity in the CSL, reflecting the Chern number of the occupied spinon band. Other Mott insulators are topologically trivial. In the Kondo-screened regime, the heavy Fermi liquid can exhibit an anomalous Hall response $(0<\sigma_{\alpha,H}<1)$, which is mostly concentrated near the CSL, as indicated by the dashed line.
(c) Lattice translation symmetry breaking. Color coding quantifies the strength of translation-symmetry breaking in bond orders. Blue corresponds to the $1\times1$ pattern, while red indicates either $2\times1$ or $2\times2$ patterns. $\Psi^{\mathrm{bdw}}$ is capped at 1 for clarity.
(d) $C_3$ rotational symmetry breaking. Color coding quantifies the strength of $C_3$ symmetry breaking in bond strengths. See main text for the definition of $\Psi^{\mathrm{bdw}}$ and $\Psi^{C_3}$.}
\label{fig:phase_diagrm}
\end{figure*}

\textit{Model Hamiltonian.}-
Motivated by these considerations, we introduce below a minimal Anderson lattice Hamiltonian for the proposed multilayer settings. Closely related lattice models have been derived for TMD trilayers from continuum moir\'e band structures combined with self-consistent Hartree-Fock calculations \cite{Kuma_PRB_Kondo}. The Hamiltonian below can be understood as a mirror-symmetric two-copy extension of the trilayer building block, which is what we mean above by a generalization of the double quantum dot model. The model Hamiltonian can be written as
\beq
H&=& -t_c  \sum_{\langle ij\rangle,\sigma,\ell} \left(c^{\dagger}_{i,\sigma,\ell}c_{j,\sigma,\ell}+ \tn{h.c.}\right) -t_f  \sum_{\langle ij \rangle,\sigma,\ell} \left(f^{\dagger}_{i,\sigma,\ell}f_{j,\sigma,\ell}+ \tn{h.c.}\right) \nonumber \\
&-&t_{\perp}  \sum_{i,\sigma,\ell} \left(c^{\dagger}_{i,\sigma,\ell}f_{i,\sigma,\ell}+ \tn{h.c.}\right) +\frac{\Delta}{2} (N^c -N^f)\nonumber\\ 
&+& \frac{U}{2} \sum_{i,\ell} n_{i,\ell}^f(n_{i,\ell}^f-1)+U' \sum_i n_{i,t}^f n_{i,b}^f. 
\eeq 
Here $\sigma\in\{\uparrow,\downarrow\}$ and $\ell\in\{t,b\}$ denote spin and layer indices, respectively. $i,j$ denote sites on the triangular lattice shared by the four layers. $n_{i,\ell}^f = n_{i,\ell,\uparrow}^f+n_{i,\ell,\downarrow}^f$ is the $f$-fermion occupation at site $i$ in the layer $\ell$. $\Delta$ represents the band offset between the $f$- and $c$-layers, which can be tuned by an external electric field. $U=e^2/(\varepsilon a)$ and $U'=e^2 / (\varepsilon\sqrt{a^2+d^2})$ are the onsite and interlayer Coulomb repulsions, respectively, where $a$ is the moir\'e lattice constant, $d$ is the layer separation between the two $f$-layers, and $\varepsilon$ is dielectric constant. We assume that Coulomb interaction on the $c$-layers is negligible. We consider the $d\ll a$ limit where the two $f$-layers are Coulomb-coupled. Introducing a single index $\alpha=1,\cdots,4 $ combining the spin and layer degrees of freedom, we obtain the following Hamiltonian with an approximate SU(4) symmetry:
\beq
\Tilde{H}&=& -t_c  \sum_{\langle ij \rangle,\alpha} \left(c^{\dagger}_{i,\alpha}c_{j,\alpha}+ \tn{h.c.}\right)-t_f  \sum_{\langle ij \rangle,\alpha} \left(f^{\dagger}_{i,\alpha}f_{j,\alpha}+ \tn{h.c.}\right) \nonumber \\
&-& t_{\perp}  \sum_{i,\alpha} \left(c^{\dagger}_{i,\alpha}f_{i,\alpha}+ \tn{h.c.}\right) + \frac{\Delta}{2}(N^c - N^f)\nonumber \\
&+& \frac{U}{2}\sum_i n_i^f (n_i^f -1)  .
\label{eq:SU4Anderson}
\eeq 

We focus on the strongly interacting regime ($U\gg |t|$) where the Coulomb-coupled $f$-layers are singly occupied ($\nu_f=1$), which corresponds to quarter filling. Upon the Schrieffer-Wolff transformation up to the third order of kinetic terms \cite{SW_prl,MacDonald_tU}, we obtain an effective SU(4) Kondo lattice Hamiltonian \cite{si}
\beq 
\bar{H}&=&T_{cc}+\sum_{i,\alpha,\beta} t_{\perp}^2 \left( \frac{c^\dagger_{i,\beta}f_{i,\beta}f^\dagger_{i,\alpha}c_{i,\alpha}}{\Delta-U} - \frac{f^\dagger_{i,\beta}c_{i,\beta}c^\dagger_{i,\alpha}f_{i,\alpha}}{\Delta}  \right) \nonumber\\
&-&\frac{2t^2_f}{U}\sum_{\langle ij \rangle,\alpha,\beta} f^\dagger_{i,\beta}f_{j,\beta}f^\dagger_{j,\alpha}f_{i,\alpha} \nonumber \\
&-&\frac{6t_f^3}{U^2}\sum_{\substack{\langle ijk \rangle \in \triangle/\triangledown \\ \alpha,\beta,\gamma}}(f^\dagger_{i,\gamma}f_{k,\gamma}f^\dagger_{k,\beta}f_{j,\beta}f^\dagger_{j,\alpha}f_{i,\alpha} + \tn{h.c.}),
\label{eq:SU4Kondo}
\eeq 
where $T_{cc}=-t_c  \sum_{\langle ij \rangle,\alpha} \left(c^{\dagger}_{i,\alpha}c_{j,\alpha}+ \tn{h.c.}\right)$ denotes the kinetic term of the $c$-fermions. The second, third, and fourth terms correspond to Kondo hybridization, two-site exchange, and three-site exchange \cite{Motrunich_PRB}, respectively. In the following, we will focus on hole doping $(t<0)$ since the low-energy bands of TMD moir\'e systems originate from monolayer valence bands.

\textit{Parton mean-field analysis.}-
We proceed by considering the mean field Hamiltonian  
\beq 
\bar{H}_{mf}&=&T_{cc}-\mu_c N^c -\sum_{i,\alpha}J_K (\Phi_i f^{\dagger}_{i,\alpha}c_{i,\alpha}+\tn{h.c.})\nonumber \\
&-& \sum_{\langle ij\rangle , \alpha} (\chi_{ij} f^\dagger_{i,\alpha}f_{j,\alpha}+\tn{h.c.} )-\sum_i \mu_i^f n^f_i,
\eeq 
with variational parameters 
\beq 
\chi_{ij}&=&J_2 A_{ji}+J_3 \sum_{\langle ijk \rangle \in \triangle/\triangledown} A_{jk}A_{ki}, \nonumber\\
A_{ij}&=&\sum_\alpha \langle f^\dagger_{i,\alpha}f_{j,\alpha}\rangle=A_{ji}^*, \nonumber\\
\Phi_i &=& \sum_\alpha \langle c^\dagger_{i,\alpha}f_{i,\alpha}\rangle.
\eeq 
Here, $J_2=2t^2_f/U$, $J_3=6t^3_f/U^2$, and $J_K=t^2_\perp(1/(U-\Delta)+1/\Delta)$ represent the two-site, three-site, and Kondo exchanges, respectively. $\mu_c$ is the chemical potential that tunes the $c$-hole filling. $\{\mu^f_i\}$ are local chemical potentials for $f$-fermions that ensure local filling constraints $\langle n^f_i\rangle=1$. We exploit various unit cells $(1\times1,\: 1\times2, \: 2\times 2)$ for variational parameters to capture possible spontaneous breaking of lattice translational symmetry. For each unit cell, we initialize $100$ random ansatzes for the variational parameters. We perform self-consistent calculations until the mean-field energy converges, and select the solution with the lowest energy. 

%\beq
%\sigma_{\alpha,H}=\sum_n\int_{1\tn{BZ}} \frac{d^2 k}{2\pi} \Omega_{n,\alpha}(\vec{k})\Theta(-E_{n,\alpha}(\vec{k})). 
%\eeq 
%Here, $\Omega_{n,\alpha}(\vec{k})$ ($E_{n,\alpha}(\vec{k})$) denotes the Berry curvature (dispersion) of the $n$-th band with flavor $\alpha\in\{1,\cdots,4\}$. Each band exhibits a mixed character of $f$- and $c$-fermions, depending on their hybridization.

To identify
different phases, we use a variety of quantitative diagnostics. Topological properties of the system are captured by the flavor Hall conductivity, $\sigma_{\alpha,H}$, which characterizes the topology of the $\alpha$-flavor band. The onset of Kondo hybridization is captured by the hybridization order parameter, 
$\Phi\equiv \frac{1}{N_{\text{sites}}}\sum_{\vr} J_K |\Phi_{\vr}|$. A finite value of $\Phi$ indicates that the $f$-fermions gain electric charge and can contribute to the electrical response of the resulting heavy Fermi liquid. In this case, $\sigma_{\alpha,H}$ is identified with the electrical Hall conductivity of the hybridized band (in units of $e^2/h$). The quantum phases we find feature various bond density wave (BDW) orders, which we capture by
\beq 
\psi^{\tn{bdw}}(\vec{Q})=\frac{1}{N_{\text{sites}}}\sum_{\langle\vr,\vr'\rangle}\left(\chi_{\vr,\vr'}e^{i\vec{Q}\cdot (\vr+\vr')/2} +\tn{c.c.} \right),
\eeq 
where $\vec{Q}$ is a momentum (within the moir\'e Brillouin zone) associated with the BDW order.

%\section{Results}
In Fig.~\ref{fig:phase_diagrm}, we present a mean-field phase diagram as a function of $J_3/J_2\:(<0)$ and $\nu_c$, with other parameters $(t_c, J_2,J_K)$ being fixed. The phase diagram exhibits a density-tuned Kondo transition at critical doping $\nu_c$, which depends on $J_3/J_2$. In the Kondo-unscreened regime ($\Phi\rightarrow 0$), the doped $c$-holes form a homogeneous Fermi liquid with a small Fermi surface, while the $f$-fermions exhibit various Mott insulating phases: a plaquette-order phase, a chiral spin liquid (CSL), and a strongly anisotropic, quasi-1D (``decoupled-chain'', DC) regime. These phases are in good qualitative agreement with a previous DMRG study of the SU(4) Hubbard model with ring exchange \cite{zhang_prl}, which also reports consistency with large-$N$ mean-field studies. In the Kondo-screened regime ($\Phi>0$), a heavy Fermi liquid appears across a wide range of $\nu_c$, and exhibits various lattice symmetry-breaking patterns. Below, we discuss the nature of each phase in detail, beginning with the Kondo-unscreened phases. We note that, as a saddle-point treatment, the phase boundaries may shift beyond mean-field, especially in gapless regimes.

%\subsection{Kondo-unscreened phases}

\begin{figure}[t]
\centering
\includegraphics[width=\linewidth]{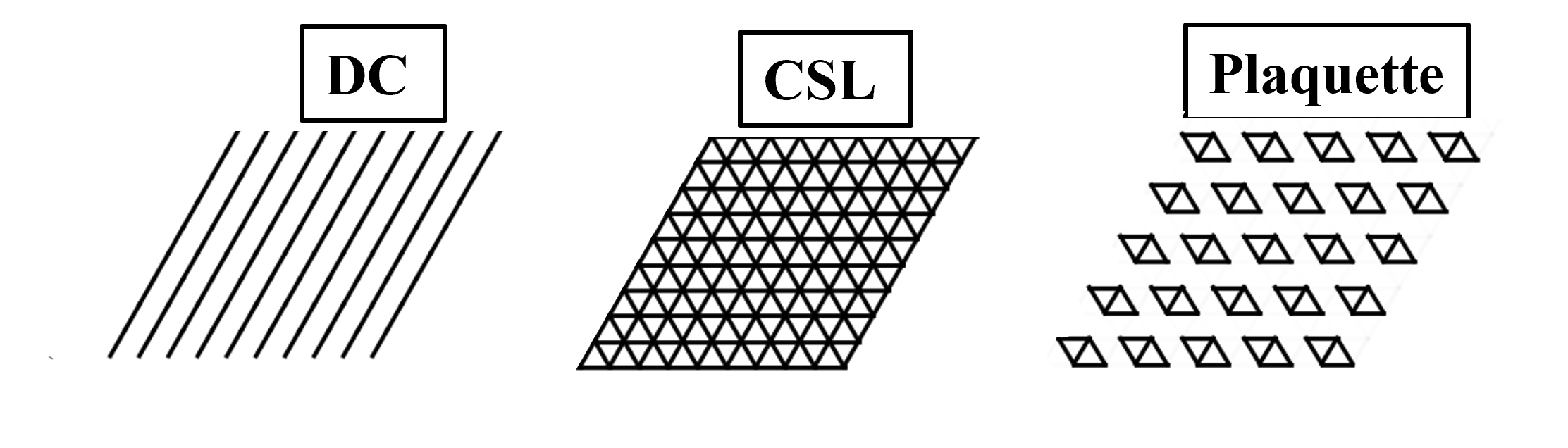} %0.9
\caption{\textbf{Real space configuration of bond strengths $|\chi_{ij}|$ in the Kondo-unscreened regime}}
\label{fig:realspace_bonds}
\end{figure}

\textit{Plaquette order.}-
When $-0.05\lesssim J_3/J_2\leq 0$, we find that the bond strength $|\chi_{ij}|$ of the Mott insulator exhibits strong lattice translational symmetry breaking, forming a $2\times 2$ plaquette order, as shown in Fig.~\ref{fig:realspace_bonds}. To understand this behavior, we consider the $J_3\rightarrow0$ limit, where the $f$-fermion Hamiltonian reduces to a tight-binding model on a triangular lattice with nearest-neighbor hopping $\chi_{ij}=J_2 A_{ji}$. Assuming a homogeneous variational solution, $A_{ji}=A_0$, the Fermi surface at quarter filling is circular for $A_0>0$, and perfectly nested for $A_0<0$. The nested case is unstable to particle-hole fluctuations at the three M points of the hexagonal moir\'e Brillouin zone, leading to a gap opening \cite{si}. Therefore, the plaquette order with a $2\times2$ unit cell is energetically favored. 

\textit{CSL.}- 
Over a finite window, $-0.25\lesssim J_3/J_2 < -0.05$, we find that the Mott insulator exhibits a quantized flavor Hall conductivity $\sigma_{\alpha,H}=1$, as shown in Fig.~\ref{fig:phase_diagrm}(b), indicating a broken time-reversal symmetry and a finite gap for neutral excitations. Such a  Mott insulator is an exotic CSL \cite{Laughlin_CSL,Wen_CSL}. To understand why $J_3<0$ stabilizes the CSL, we consider the three-site term acting on a triangular plaquette
\beq 
H_\triangle &=& -J_3 (P_{123}+\tn{h.c.}) ,
\eeq 
where $P_{123}=|\alpha_1 \beta_2 \gamma_3\rangle \langle \gamma_1 \alpha_2 \beta_3|$ is a ring exchange operator. Let us denote cyclic eigenvalues of $P_{123}$ by $\lambda \in \{1,\omega,\omega^2 \}$ with $\omega=e^{i2\pi/3}$. The eigenvalues of the three-site term are given by $-J_3 (\lambda+\lambda^{-1})$, indicating two-fold degenerate ground states with energy $E=-J_3(\omega+\omega^{-1})=J_3<0$. The two ground states have opposite chirality $\chi\equiv (P_{123}-\tn{h.c.})/2i = \pm \frac{\sqrt{3}}{2}$, implying that the ground state spontaneously breaks time-reversal symmetry. The ground state exhibits a staggered flux $\sim\pi/2$ per unit cell $(\triucell)$. The staggered fluxes sum to $2\pi$ over an enlarged $2\times2$ unit cell, and therefore each $f$-band splits into four sub-bands. The CSL can thus be viewed as a Chern insulator of neutral \textit{spinons}, where for each flavor the lowest-energy Chern band with $C=1$ is occupied. The bond strengths $|\chi_{ij}|$ are nearly homogeneous. It was recently proposed that doping the SU(4) CSL can lead to an exotic superconductor \cite{Zhang_Song_SU4SC}.

\textit{DC.}-
When $J_3$ is decreased to more negative values, the Mott insulator forms a DC phase, breaking $C_3$ rotational symmetry of the bond strengths, as shown in Fig.~\ref{fig:realspace_bonds}. To understand this behavior, we introduce the generators $\{T^a\}\:(a=1,\cdots,15)$ of the fundamental representation of SU(4), which satisfy the anticommutation relation $\{T^a,T^b\}=\frac{1}{4}\delta_{ab}+d_{abc}T^c$, where $d_{abc}$ is the symmetric tensor. The three-site term can then be rewritten as a sum of two-site and three-site couplings \cite{si}, 
\beq 
H_3 = -2J_3\sum_{\langle ij \rangle} T^a_i T^a_j - 4 J_3 \sum_{\langle ijk \rangle} d_{abc} T^a_i T^b_j T^c_k+\tn{(const)}.
\eeq 
For sufficiently negative $J_3$, the two-site coupling drives a strong bond ordering, leading to the DC phase. We note that the three-site coupling term ($\propto d_{abc}$) survives in the SU($N>2$) case even in the absence of an external magnetic flux, in contrast to the SU(2) case. 

\begin{figure*}[tbh]
\centering
\includegraphics[width=\linewidth]{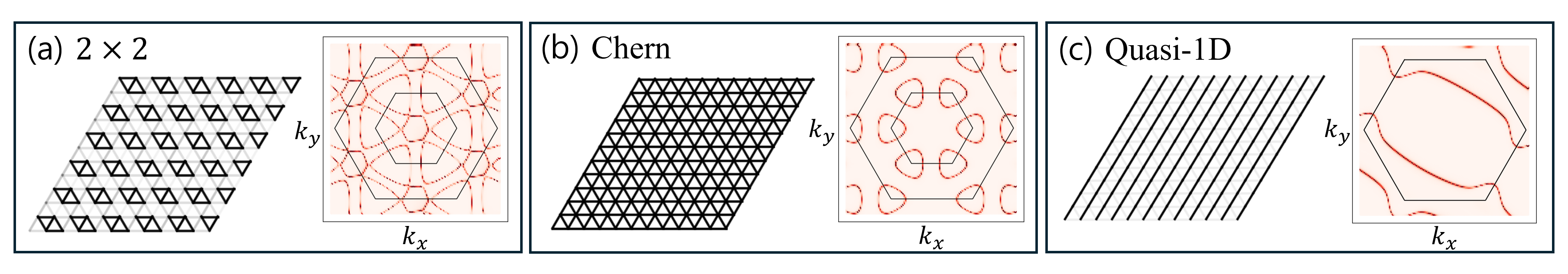} %0.9
\caption{\textbf{HFLs with different symmetry breaking patterns}. Bond strengths $(|\chi_{ij}|)$ and fermiology at representative parameter sets are shown. (a) $2\times 2$ CDW metal at $(J_3,\nu_c)=(0,1.47)$. (b) Chern metal at $(J_3,\nu_c)=(-0.2,0.38)$. (c) Quasi-1D metal at $(J_3,\nu_c)=(-0.4,0.33)$. Black hexagons represent the original and reduced Brillouin zones.}
\label{fig:HFL}
\end{figure*}

\textit{Heavy Fermi liquids.}-
With increasing $\nu_c$, Kondo hybridization sets in, giving rise to heavy Fermi liquids (HFLs). We find that the hybridized phase becomes a compensated semimetal at the commensurate filling $\nu_c=1$ and remains gapless. The BDW order of the underlying Mott insulator in the Kondo-unscreened regime is imprinted onto proximate HFLs, whereas at higher doping the HFL becomes homogeneous. The BDW order breaks lattice translation symmetry and/or $C_3$ rotational symmetry. To quantify translation symmetry breaking, we define
\beq
\Psi^{\mathrm{bdw}} = \frac{\max{\psi^{\mathrm{bdw}}(M_i)}}{\psi^{\mathrm{bdw}}(\Gamma)},
\eeq
where $\Gamma=(0,0)$ and $M_{i=1,2,3}$ denote three M points in the moir\'e Brillouin zone. $\Psi^{\mathrm{bdw}}\sim O(1)$ indicates that the bond order develops an enlarged unit cell ($2\times1$ or $2\times2$). To quantify $C_3$ rotational symmetry breaking of the bond strength, we use 
\beq
\Psi^{C_3}=\frac{1}{N_{\text{sites}}}\sum_{\vr}
\frac{\mathrm{std}\left(|\chi_{\vr,\vr+\ve_\tau}|\right)}{\mathrm{mean}\left(|\chi_{\vr,\vr+\ve_\tau}|\right)},
\eeq
where $\ve_{1}=(1,0),\:\ve_{2}=\left(-\frac{1}{2},\frac{\sqrt{3}}{2}\right),\ve_{3}=\left(-\frac{1}{2},-\frac{\sqrt{3}}{2}\right)$ are three lattice vectors (lattice constant set to unity). Here std and mean denote the standard deviation and average, respectively, taken over the three nearest-neighbor bonds of site $\vr$. The results of these diagnostics are shown in Fig.~\ref{fig:phase_diagrm}(c,d). Combining these measures with the real-space configuration of $|\chi_{ij}|$ and the fermiology (Fig.~\ref{fig:HFL}), we identify three distinct HFLs with different symmetry-breaking patterns.

%\textit{$2\times 2$ CDW metal.}-
We first note that the Plaquette phase exhibits a much larger critical $\nu_c$ for Kondo transition than that of the other Kondo-unscreened phases, as shown in Fig.~\ref{fig:phase_diagrm}(a). This can be understood from the fact that, in the plaquette phase, four spins within a plaquette form a (nearly) SU(4)-singlet state \cite{Arovas_SUNsinglet,Hermele_SU4plaquette,Xu_SU4Plaquette}, which is energetically stable and requires higher doping to be broken by Kondo hybridization. Above critical $\nu_c$, a $2\times 2$ CDW metal with broken $C_3$ rotational symmetry emerges with bond orders that retain the structure of weakly-coupled plaquettes, as shown in Fig.~\ref{fig:HFL}(a). At the mean-field level, the Kondo transition is first order, as evidenced by a sudden jump in $\nu_c$.

%\textit{Chern metal.}-
Adjacent to the CSL, we find an HFL with an anomalous Hall response, $0<\sigma_{\alpha,H}<1$, which we dub a Chern metal. The Chern metal inherits the nontrivial topology of the CSL \cite{Kuma_PRB_Kondo} and, due to the underlying flux pattern, features a $2\times 2$ unit cell, as shown in Fig.~\ref{fig:HFL}(b). Recently, experimental evidence of a Chern metal was reported in MoTe$_2$/WSe$_2$ heterobilayers near the Kondo breakdown transition \cite{zhao_chernmetal}, suggesting the interesting possibility that local moments in the MoTe$_2$ layer form an exotic Mott insulator below the critical doping. 

%\textit{Quasi-1D metal.}-
The HFL proximate to the DC phase features strong $C_3$-symmetry breaking, with bond orders forming weakly-coupled chains. Consequently, the Fermi surface exhibits a quasi-1D character, as shown in Fig.~\ref{fig:HFL}(c). A strong anisotropy in the electrical resistivity would serve as a clear experimental signature. 

%Our mean-field calculations show that the CDW order is described either by a stripy pattern or a $2\times 2$ unit cell. The stripy pattern is characterized by a prominent peak at a single M-point, while the $2\times 2$ pattern exhibits comparable peaks at multiple M-points. To quantitatively distinguish these patterns, we introduce a dimensionless parameter
%\beq 
%x=\frac{\tn{Max} (I_1,I_2,I_3)  }{\tn{Median}(I_1,I_2,I_3)},
%\eeq 
%with $I_i=\Psi^{\tn{cdw}}(\vQ_i)$. A stripy CDW pattern is signaled by $x$ deviating significantly from 1. The magnitude $(\sum_{i=1,2,3}I_i)$ and symmetry breaking pattern $(x)$ of the CDW order are shown in Fig.~\ref{fig:phase_diagrm}(c). Larger blobs represent stronger CDW order, while color distinguishes the CDW pattern: red indicates a stripy pattern, and blue represents a $2\times 2$ pattern. For color coding, $x$ is capped at 100 for better visibility. As indicated by the colors, the CDW pattern largely follows the BDW pattern of the proximate Mott insulator; see Fig.XXX for the real-space configurations of $\Phi_i$ and $n^c_i$ in the stripy CDW metal near the DC. Moreover, the CDW order becomes weaker at larger doping, as stronger hybridization suppresses the tendency toward lattice translational symmetry breaking inherited from the insulating ground state. As a result, the symmetry breaking is quenched near the total commensurate filling $\nu_T=2$, leading to partially filled bands and the absence of Kondo insulating phases.   

%we leave a detailed analysis of such effects for future work

\textit{Outlook.}-
In this work, we proposed TMD multilayer settings that realize an SU(4) Kondo lattice and performed a self-consistent mean-field study of its phase diagram. We have focused on a specific choice of parameters that does not necessarily derive from a microscopic model, since our main goal is to outline possible quantum phases that can be realized in this setting. An interesting future direction is to incorporate realistic microscopic parameters (e.g., extracted from continuum moir\'e models and self-consistent Hartree-Fock treatments of TMD multilayers) to map out the phase diagram as a function of gate voltages, including the effects of RKKY interactions, and to study the robustness of each phase against SU(4)-breaking couplings \cite{Trebst_brokenSU4}.
 It will also be important to carefully consider layer hybridizations in order to ensure the emergence of well-defined $f$- and $c$-layers \cite{ShiZeng_KI}. It would also be interesting to work with fully projected (Gutzwiller-projected) wavefunctions, which can quantitatively renormalize energies and shift mean-field phase boundaries, with potentially stronger effects in gapless regimes such as the heavy Fermi liquid.

As noted earlier, MoTe$_2$/WSe$_2$ heterobilayers could also serve as building blocks for Kondo lattices. A microscopic description of AB-stacked MoTe$_2$/WSe$_2$ reveals spin-dependent hopping phases \cite{QAH_DFT_zhang_2021,QAH_devakul_2022,Rademaker_hetero_2022,Guerci_2023}, which break spin-rotational symmetry and thus prevent full SU(4) symmetry. Nevertheless, such heterobilayer-based settings may host additional topological phases due to the chiral nature of hybridization \cite{Guerci_2023}, as well as additional virtual hopping processes that are allowed due to the honeycomb geometry \cite{si}. We leave a detailed analysis of this case for future work.

Another promising direction is to study instabilities of the HFLs. For instance, the Chern metal hosts hole-like Fermi pockets with finite Berry curvature, which may give rise to instabilities in the particle-particle channel that can lead to unconventional superconductivity \cite{Chern_FP_PDW}. 

Several experimental challenges need to be addressed to ensure tunability of the proposed setting. Realizing the SU(4)-symmetric model requires maintaining layer symmetry between $f$-layers and $c$-layers, which can be achieved by applying the displacement field in a mirror-symmetric fashion. This in principle requires three gates: top, bottom, and middle. Developing a practical method to achieve this symmetry will be crucial.

{\it Acknowledgments.-} 
I thank T. Senthil, Allan MacDonald, and Jie Shan for stimulating discussions at the Boulder Summer School 2025. I am grateful to Kin Fai Mak and Wenjin Zhao for collaboration on a related project, and especially to Kin Fai Mak for insightful discussions during my A exam that helped motivate this work. I thank Aaditya Panigrahi, Haoyu Guo, Omri Lesser, and Xuepeng Wang for helpful discussions and comments. I am deeply indebted to Debanjan Chowdhury for his guidance throughout my PhD and invaluable discussions related to this work. I acknowledge the support of the group and access to shared computational resources.

\bibliography{refs}

\begin{thebibliography}{73}%
\makeatletter
\providecommand \@ifxundefined [1]{%
 \@ifx{#1\undefined}
}%
\providecommand \@ifnum [1]{%
 \ifnum #1\expandafter \@firstoftwo
 \else \expandafter \@secondoftwo
 \fi
}%
\providecommand \@ifx [1]{%
 \ifx #1\expandafter \@firstoftwo
 \else \expandafter \@secondoftwo
 \fi
}%
\providecommand \natexlab [1]{#1}%
\providecommand \enquote  [1]{``#1''}%
\providecommand \bibnamefont  [1]{#1}%
\providecommand \bibfnamefont [1]{#1}%
\providecommand \citenamefont [1]{#1}%
\providecommand \href@noop [0]{\@secondoftwo}%
\providecommand \href [0]{\begingroup \@sanitize@url \@href}%
\providecommand \@href[1]{\@@startlink{#1}\@@href}%
\providecommand \@@href[1]{\endgroup#1\@@endlink}%
\providecommand \@sanitize@url [0]{\catcode `\\12\catcode `\$12\catcode `\&12\catcode `\#12\catcode `\^12\catcode `\_12\catcode `\%12\relax}%
\providecommand \@@startlink[1]{}%
\providecommand \@@endlink[0]{}%
\providecommand \url  [0]{\begingroup\@sanitize@url \@url }%
\providecommand \@url [1]{\endgroup\@href {#1}{\urlprefix }}%
\providecommand \urlprefix  [0]{URL }%
\providecommand \Eprint [0]{\href }%
\providecommand \doibase [0]{https://doi.org/}%
\providecommand \selectlanguage [0]{\@gobble}%
\providecommand \bibinfo  [0]{\@secondoftwo}%
\providecommand \bibfield  [0]{\@secondoftwo}%
\providecommand \translation [1]{[#1]}%
\providecommand \BibitemOpen [0]{}%
\providecommand \bibitemStop [0]{}%
\providecommand \bibitemNoStop [0]{.\EOS\space}%
\providecommand \EOS [0]{\spacefactor3000\relax}%
\providecommand \BibitemShut  [1]{\csname bibitem#1\endcsname}%
\let\auto@bib@innerbib\@empty
%</preamble>
\bibitem [{\citenamefont {Wu}\ \emph {et~al.}(2018)\citenamefont {Wu}, \citenamefont {Lovorn}, \citenamefont {Tutuc},\ and\ \citenamefont {MacDonald}}]{Wu_Hubbard_PRL}%
  \BibitemOpen
  \bibfield  {author} {\bibinfo {author} {\bibfnamefont {F.}~\bibnamefont {Wu}}, \bibinfo {author} {\bibfnamefont {T.}~\bibnamefont {Lovorn}}, \bibinfo {author} {\bibfnamefont {E.}~\bibnamefont {Tutuc}},\ and\ \bibinfo {author} {\bibfnamefont {A.~H.}\ \bibnamefont {MacDonald}},\ }\bibfield  {title} {\bibinfo {title} {Hubbard model physics in transition metal dichalcogenide moir\'e bands},\ }\href {https://doi.org/10.1103/PhysRevLett.121.026402} {\bibfield  {journal} {\bibinfo  {journal} {Phys. Rev. Lett.}\ }\textbf {\bibinfo {volume} {121}},\ \bibinfo {pages} {026402} (\bibinfo {year} {2018})}\BibitemShut {NoStop}%
\bibitem [{\citenamefont {Wu}\ \emph {et~al.}(2019)\citenamefont {Wu}, \citenamefont {Lovorn}, \citenamefont {Tutuc}, \citenamefont {Martin},\ and\ \citenamefont {MacDonald}}]{Wu_QSH_PRL}%
  \BibitemOpen
  \bibfield  {author} {\bibinfo {author} {\bibfnamefont {F.}~\bibnamefont {Wu}}, \bibinfo {author} {\bibfnamefont {T.}~\bibnamefont {Lovorn}}, \bibinfo {author} {\bibfnamefont {E.}~\bibnamefont {Tutuc}}, \bibinfo {author} {\bibfnamefont {I.}~\bibnamefont {Martin}},\ and\ \bibinfo {author} {\bibfnamefont {A.~H.}\ \bibnamefont {MacDonald}},\ }\bibfield  {title} {\bibinfo {title} {Topological insulators in twisted transition metal dichalcogenide homobilayers},\ }\href {https://doi.org/10.1103/PhysRevLett.122.086402} {\bibfield  {journal} {\bibinfo  {journal} {Phys. Rev. Lett.}\ }\textbf {\bibinfo {volume} {122}},\ \bibinfo {pages} {086402} (\bibinfo {year} {2019})}\BibitemShut {NoStop}%
\bibitem [{\citenamefont {Mak}\ and\ \citenamefont {Shan}(2022)}]{Mak_Shan_review}%
  \BibitemOpen
  \bibfield  {author} {\bibinfo {author} {\bibfnamefont {K.~F.}\ \bibnamefont {Mak}}\ and\ \bibinfo {author} {\bibfnamefont {J.}~\bibnamefont {Shan}},\ }\bibfield  {title} {\bibinfo {title} {Semiconductor moir\'e materials},\ }\href {https://doi.org/10.1038/s41565-022-01165-6} {\bibfield  {journal} {\bibinfo  {journal} {Nature Nanotechnology}\ }\textbf {\bibinfo {volume} {17}},\ \bibinfo {pages} {686–695} (\bibinfo {year} {2022})}\BibitemShut {NoStop}%
\bibitem [{\citenamefont {Tang}\ \emph {et~al.}(2020)\citenamefont {Tang}, \citenamefont {Li}, \citenamefont {Li}, \citenamefont {Xu}, \citenamefont {Liu}, \citenamefont {Barmak}, \citenamefont {Watanabe}, \citenamefont {Taniguchi}, \citenamefont {MacDonald}, \citenamefont {Shan},\ and\ \citenamefont {Mak}}]{Tang_simulation_2020}%
  \BibitemOpen
  \bibfield  {author} {\bibinfo {author} {\bibfnamefont {Y.}~\bibnamefont {Tang}}, \bibinfo {author} {\bibfnamefont {L.}~\bibnamefont {Li}}, \bibinfo {author} {\bibfnamefont {T.}~\bibnamefont {Li}}, \bibinfo {author} {\bibfnamefont {Y.}~\bibnamefont {Xu}}, \bibinfo {author} {\bibfnamefont {S.}~\bibnamefont {Liu}}, \bibinfo {author} {\bibfnamefont {K.}~\bibnamefont {Barmak}}, \bibinfo {author} {\bibfnamefont {K.}~\bibnamefont {Watanabe}}, \bibinfo {author} {\bibfnamefont {T.}~\bibnamefont {Taniguchi}}, \bibinfo {author} {\bibfnamefont {A.~H.}\ \bibnamefont {MacDonald}}, \bibinfo {author} {\bibfnamefont {J.}~\bibnamefont {Shan}},\ and\ \bibinfo {author} {\bibfnamefont {K.~F.}\ \bibnamefont {Mak}},\ }\bibfield  {title} {\bibinfo {title} {Simulation of hubbard model physics in wse2/ws2 moir\'e superlattices},\ }\href {https://doi.org/10.1038/s41586-020-2085-3} {\bibfield  {journal} {\bibinfo  {journal} {Nature}\ }\textbf {\bibinfo {volume} {579}},\ \bibinfo {pages} {353–358} (\bibinfo {year}
  {2020})}\BibitemShut {NoStop}%
\bibitem [{\citenamefont {Regan}\ \emph {et~al.}(2020)\citenamefont {Regan}, \citenamefont {Wang}, \citenamefont {Jin}, \citenamefont {Bakti~Utama}, \citenamefont {Gao}, \citenamefont {Wei}, \citenamefont {Zhao}, \citenamefont {Zhao}, \citenamefont {Zhang}, \citenamefont {Yumigeta}, \citenamefont {Blei}, \citenamefont {Carlström}, \citenamefont {Watanabe}, \citenamefont {Taniguchi}, \citenamefont {Tongay}, \citenamefont {Crommie}, \citenamefont {Zettl},\ and\ \citenamefont {Wang}}]{Regan_mottwigner_2020}%
  \BibitemOpen
  \bibfield  {author} {\bibinfo {author} {\bibfnamefont {E.~C.}\ \bibnamefont {Regan}}, \bibinfo {author} {\bibfnamefont {D.}~\bibnamefont {Wang}}, \bibinfo {author} {\bibfnamefont {C.}~\bibnamefont {Jin}}, \bibinfo {author} {\bibfnamefont {M.~I.}\ \bibnamefont {Bakti~Utama}}, \bibinfo {author} {\bibfnamefont {B.}~\bibnamefont {Gao}}, \bibinfo {author} {\bibfnamefont {X.}~\bibnamefont {Wei}}, \bibinfo {author} {\bibfnamefont {S.}~\bibnamefont {Zhao}}, \bibinfo {author} {\bibfnamefont {W.}~\bibnamefont {Zhao}}, \bibinfo {author} {\bibfnamefont {Z.}~\bibnamefont {Zhang}}, \bibinfo {author} {\bibfnamefont {K.}~\bibnamefont {Yumigeta}}, \bibinfo {author} {\bibfnamefont {M.}~\bibnamefont {Blei}}, \bibinfo {author} {\bibfnamefont {J.~D.}\ \bibnamefont {Carlström}}, \bibinfo {author} {\bibfnamefont {K.}~\bibnamefont {Watanabe}}, \bibinfo {author} {\bibfnamefont {T.}~\bibnamefont {Taniguchi}}, \bibinfo {author} {\bibfnamefont {S.}~\bibnamefont {Tongay}}, \bibinfo {author} {\bibfnamefont {M.}~\bibnamefont {Crommie}},
  \bibinfo {author} {\bibfnamefont {A.}~\bibnamefont {Zettl}},\ and\ \bibinfo {author} {\bibfnamefont {F.}~\bibnamefont {Wang}},\ }\bibfield  {title} {\bibinfo {title} {Mott and generalized wigner crystal states in wse2/ws2 moir\'e superlattices},\ }\href {https://doi.org/10.1038/s41586-020-2092-4} {\bibfield  {journal} {\bibinfo  {journal} {Nature}\ }\textbf {\bibinfo {volume} {579}},\ \bibinfo {pages} {359–363} (\bibinfo {year} {2020})}\BibitemShut {NoStop}%
\bibitem [{\citenamefont {Xu}\ \emph {et~al.}(2020)\citenamefont {Xu}, \citenamefont {Liu}, \citenamefont {Rhodes}, \citenamefont {Watanabe}, \citenamefont {Taniguchi}, \citenamefont {Hone}, \citenamefont {Elser}, \citenamefont {Mak},\ and\ \citenamefont {Shan}}]{Xu_mottwigner_2020}%
  \BibitemOpen
  \bibfield  {author} {\bibinfo {author} {\bibfnamefont {Y.}~\bibnamefont {Xu}}, \bibinfo {author} {\bibfnamefont {S.}~\bibnamefont {Liu}}, \bibinfo {author} {\bibfnamefont {D.~A.}\ \bibnamefont {Rhodes}}, \bibinfo {author} {\bibfnamefont {K.}~\bibnamefont {Watanabe}}, \bibinfo {author} {\bibfnamefont {T.}~\bibnamefont {Taniguchi}}, \bibinfo {author} {\bibfnamefont {J.}~\bibnamefont {Hone}}, \bibinfo {author} {\bibfnamefont {V.}~\bibnamefont {Elser}}, \bibinfo {author} {\bibfnamefont {K.~F.}\ \bibnamefont {Mak}},\ and\ \bibinfo {author} {\bibfnamefont {J.}~\bibnamefont {Shan}},\ }\bibfield  {title} {\bibinfo {title} {Correlated insulating states at fractional fillings of moir\'e superlattices},\ }\href {https://doi.org/10.1038/s41586-020-2868-6} {\bibfield  {journal} {\bibinfo  {journal} {Nature}\ }\textbf {\bibinfo {volume} {587}},\ \bibinfo {pages} {214–218} (\bibinfo {year} {2020})}\BibitemShut {NoStop}%
\bibitem [{\citenamefont {Li}\ \emph {et~al.}(2021{\natexlab{a}})\citenamefont {Li}, \citenamefont {Li}, \citenamefont {Regan}, \citenamefont {Wang}, \citenamefont {Zhao}, \citenamefont {Kahn}, \citenamefont {Yumigeta}, \citenamefont {Blei}, \citenamefont {Taniguchi}, \citenamefont {Watanabe}, \citenamefont {Tongay}, \citenamefont {Zettl}, \citenamefont {Crommie},\ and\ \citenamefont {Wang}}]{Li_imagewigner_2021}%
  \BibitemOpen
  \bibfield  {author} {\bibinfo {author} {\bibfnamefont {H.}~\bibnamefont {Li}}, \bibinfo {author} {\bibfnamefont {S.}~\bibnamefont {Li}}, \bibinfo {author} {\bibfnamefont {E.~C.}\ \bibnamefont {Regan}}, \bibinfo {author} {\bibfnamefont {D.}~\bibnamefont {Wang}}, \bibinfo {author} {\bibfnamefont {W.}~\bibnamefont {Zhao}}, \bibinfo {author} {\bibfnamefont {S.}~\bibnamefont {Kahn}}, \bibinfo {author} {\bibfnamefont {K.}~\bibnamefont {Yumigeta}}, \bibinfo {author} {\bibfnamefont {M.}~\bibnamefont {Blei}}, \bibinfo {author} {\bibfnamefont {T.}~\bibnamefont {Taniguchi}}, \bibinfo {author} {\bibfnamefont {K.}~\bibnamefont {Watanabe}}, \bibinfo {author} {\bibfnamefont {S.}~\bibnamefont {Tongay}}, \bibinfo {author} {\bibfnamefont {A.}~\bibnamefont {Zettl}}, \bibinfo {author} {\bibfnamefont {M.~F.}\ \bibnamefont {Crommie}},\ and\ \bibinfo {author} {\bibfnamefont {F.}~\bibnamefont {Wang}},\ }\bibfield  {title} {\bibinfo {title} {Imaging two-dimensional generalized wigner crystals},\ }\href
  {https://doi.org/10.1038/s41586-021-03874-9} {\bibfield  {journal} {\bibinfo  {journal} {Nature}\ }\textbf {\bibinfo {volume} {597}},\ \bibinfo {pages} {650–654} (\bibinfo {year} {2021}{\natexlab{a}})}\BibitemShut {NoStop}%
\bibitem [{\citenamefont {Li}\ \emph {et~al.}(2021{\natexlab{b}})\citenamefont {Li}, \citenamefont {Jiang}, \citenamefont {Li}, \citenamefont {Zhang}, \citenamefont {Kang}, \citenamefont {Zhu}, \citenamefont {Watanabe}, \citenamefont {Taniguchi}, \citenamefont {Chowdhury}, \citenamefont {Fu}, \citenamefont {Shan},\ and\ \citenamefont {Mak}}]{li_continuous_2021}%
  \BibitemOpen
  \bibfield  {author} {\bibinfo {author} {\bibfnamefont {T.}~\bibnamefont {Li}}, \bibinfo {author} {\bibfnamefont {S.}~\bibnamefont {Jiang}}, \bibinfo {author} {\bibfnamefont {L.}~\bibnamefont {Li}}, \bibinfo {author} {\bibfnamefont {Y.}~\bibnamefont {Zhang}}, \bibinfo {author} {\bibfnamefont {K.}~\bibnamefont {Kang}}, \bibinfo {author} {\bibfnamefont {J.}~\bibnamefont {Zhu}}, \bibinfo {author} {\bibfnamefont {K.}~\bibnamefont {Watanabe}}, \bibinfo {author} {\bibfnamefont {T.}~\bibnamefont {Taniguchi}}, \bibinfo {author} {\bibfnamefont {D.}~\bibnamefont {Chowdhury}}, \bibinfo {author} {\bibfnamefont {L.}~\bibnamefont {Fu}}, \bibinfo {author} {\bibfnamefont {J.}~\bibnamefont {Shan}},\ and\ \bibinfo {author} {\bibfnamefont {K.~F.}\ \bibnamefont {Mak}},\ }\bibfield  {title} {\bibinfo {title} {Continuous {Mott} transition in semiconductor moir\'{e} superlattices},\ }\href {https://doi.org/10.1038/s41586-021-03853-0} {\bibfield  {journal} {\bibinfo  {journal} {Nature}\ }\textbf {\bibinfo {volume} {597}},\ \bibinfo
  {pages} {350} (\bibinfo {year} {2021}{\natexlab{b}})}\BibitemShut {NoStop}%
\bibitem [{\citenamefont {Ghiotto}\ \emph {et~al.}(2021)\citenamefont {Ghiotto}, \citenamefont {Shih}, \citenamefont {Pereira}, \citenamefont {Rhodes}, \citenamefont {Kim}, \citenamefont {Zang}, \citenamefont {Millis}, \citenamefont {Watanabe}, \citenamefont {Taniguchi}, \citenamefont {Hone}, \citenamefont {Wang}, \citenamefont {Dean},\ and\ \citenamefont {Pasupathy}}]{ghiotto_quantum_2021}%
  \BibitemOpen
  \bibfield  {author} {\bibinfo {author} {\bibfnamefont {A.}~\bibnamefont {Ghiotto}}, \bibinfo {author} {\bibfnamefont {E.-M.}\ \bibnamefont {Shih}}, \bibinfo {author} {\bibfnamefont {G.~S. S.~G.}\ \bibnamefont {Pereira}}, \bibinfo {author} {\bibfnamefont {D.~A.}\ \bibnamefont {Rhodes}}, \bibinfo {author} {\bibfnamefont {B.}~\bibnamefont {Kim}}, \bibinfo {author} {\bibfnamefont {J.}~\bibnamefont {Zang}}, \bibinfo {author} {\bibfnamefont {A.~J.}\ \bibnamefont {Millis}}, \bibinfo {author} {\bibfnamefont {K.}~\bibnamefont {Watanabe}}, \bibinfo {author} {\bibfnamefont {T.}~\bibnamefont {Taniguchi}}, \bibinfo {author} {\bibfnamefont {J.~C.}\ \bibnamefont {Hone}}, \bibinfo {author} {\bibfnamefont {L.}~\bibnamefont {Wang}}, \bibinfo {author} {\bibfnamefont {C.~R.}\ \bibnamefont {Dean}},\ and\ \bibinfo {author} {\bibfnamefont {A.~N.}\ \bibnamefont {Pasupathy}},\ }\bibfield  {title} {\bibinfo {title} {Quantum criticality in twisted transition metal dichalcogenides},\ }\href {https://doi.org/10.1038/s41586-021-03815-6}
  {\bibfield  {journal} {\bibinfo  {journal} {Nature}\ }\textbf {\bibinfo {volume} {597}},\ \bibinfo {pages} {345} (\bibinfo {year} {2021})}\BibitemShut {NoStop}%
\bibitem [{\citenamefont {Savary}\ and\ \citenamefont {Balents}(2016)}]{Savary_2016}%
  \BibitemOpen
  \bibfield  {author} {\bibinfo {author} {\bibfnamefont {L.}~\bibnamefont {Savary}}\ and\ \bibinfo {author} {\bibfnamefont {L.}~\bibnamefont {Balents}},\ }\bibfield  {title} {\bibinfo {title} {Quantum spin liquids: a review},\ }\href {https://doi.org/10.1088/0034-4885/80/1/016502} {\bibfield  {journal} {\bibinfo  {journal} {Reports on Progress in Physics}\ }\textbf {\bibinfo {volume} {80}},\ \bibinfo {pages} {016502} (\bibinfo {year} {2016})}\BibitemShut {NoStop}%
\bibitem [{\citenamefont {Zhou}\ \emph {et~al.}(2017)\citenamefont {Zhou}, \citenamefont {Kanoda},\ and\ \citenamefont {Ng}}]{QSL_RMP}%
  \BibitemOpen
  \bibfield  {author} {\bibinfo {author} {\bibfnamefont {Y.}~\bibnamefont {Zhou}}, \bibinfo {author} {\bibfnamefont {K.}~\bibnamefont {Kanoda}},\ and\ \bibinfo {author} {\bibfnamefont {T.-K.}\ \bibnamefont {Ng}},\ }\bibfield  {title} {\bibinfo {title} {Quantum spin liquid states},\ }\href {https://doi.org/10.1103/RevModPhys.89.025003} {\bibfield  {journal} {\bibinfo  {journal} {Rev. Mod. Phys.}\ }\textbf {\bibinfo {volume} {89}},\ \bibinfo {pages} {025003} (\bibinfo {year} {2017})}\BibitemShut {NoStop}%
\bibitem [{\citenamefont {Broholm}\ \emph {et~al.}(2020)\citenamefont {Broholm}, \citenamefont {Cava}, \citenamefont {Kivelson}, \citenamefont {Nocera}, \citenamefont {Norman},\ and\ \citenamefont {Senthil}}]{Broholm_2020}%
  \BibitemOpen
  \bibfield  {author} {\bibinfo {author} {\bibfnamefont {C.}~\bibnamefont {Broholm}}, \bibinfo {author} {\bibfnamefont {R.~J.}\ \bibnamefont {Cava}}, \bibinfo {author} {\bibfnamefont {S.~A.}\ \bibnamefont {Kivelson}}, \bibinfo {author} {\bibfnamefont {D.~G.}\ \bibnamefont {Nocera}}, \bibinfo {author} {\bibfnamefont {M.~R.}\ \bibnamefont {Norman}},\ and\ \bibinfo {author} {\bibfnamefont {T.}~\bibnamefont {Senthil}},\ }\bibfield  {title} {\bibinfo {title} {Quantum spin liquids},\ }\bibfield  {journal} {\bibinfo  {journal} {Science}\ }\textbf {\bibinfo {volume} {367}},\ \href {https://doi.org/10.1126/science.aay0668} {10.1126/science.aay0668} (\bibinfo {year} {2020})\BibitemShut {NoStop}%
\bibitem [{\citenamefont {Doniach}(1977)}]{Doniach_1977}%
  \BibitemOpen
  \bibfield  {author} {\bibinfo {author} {\bibfnamefont {S.}~\bibnamefont {Doniach}},\ }\bibfield  {title} {\bibinfo {title} {The kondo lattice and weak antiferromagnetism},\ }\href {https://doi.org/10.1016/0378-4363(77)90190-5} {\bibfield  {journal} {\bibinfo  {journal} {Physica B+C}\ }\textbf {\bibinfo {volume} {91}},\ \bibinfo {pages} {231–234} (\bibinfo {year} {1977})}\BibitemShut {NoStop}%
\bibitem [{\citenamefont {Stewart}(1984)}]{HFL_RMP}%
  \BibitemOpen
  \bibfield  {author} {\bibinfo {author} {\bibfnamefont {G.~R.}\ \bibnamefont {Stewart}},\ }\bibfield  {title} {\bibinfo {title} {Heavy-fermion systems},\ }\href {https://doi.org/10.1103/RevModPhys.56.755} {\bibfield  {journal} {\bibinfo  {journal} {Rev. Mod. Phys.}\ }\textbf {\bibinfo {volume} {56}},\ \bibinfo {pages} {755} (\bibinfo {year} {1984})}\BibitemShut {NoStop}%
\bibitem [{\citenamefont {Coleman}\ \emph {et~al.}(2001)\citenamefont {Coleman}, \citenamefont {Pépin}, \citenamefont {Si},\ and\ \citenamefont {Ramazashvili}}]{Coleman_Si_HFLreview}%
  \BibitemOpen
  \bibfield  {author} {\bibinfo {author} {\bibfnamefont {P.}~\bibnamefont {Coleman}}, \bibinfo {author} {\bibfnamefont {C.}~\bibnamefont {Pépin}}, \bibinfo {author} {\bibfnamefont {Q.}~\bibnamefont {Si}},\ and\ \bibinfo {author} {\bibfnamefont {R.}~\bibnamefont {Ramazashvili}},\ }\bibfield  {title} {\bibinfo {title} {How do fermi liquids get heavy and die?},\ }\href {https://doi.org/10.1088/0953-8984/13/35/202} {\bibfield  {journal} {\bibinfo  {journal} {Journal of Physics: Condensed Matter}\ }\textbf {\bibinfo {volume} {13}},\ \bibinfo {pages} {R723–R738} (\bibinfo {year} {2001})}\BibitemShut {NoStop}%
\bibitem [{\citenamefont {Senthil}\ \emph {et~al.}(2003)\citenamefont {Senthil}, \citenamefont {Sachdev},\ and\ \citenamefont {Vojta}}]{Senthil_FL*_PRL}%
  \BibitemOpen
  \bibfield  {author} {\bibinfo {author} {\bibfnamefont {T.}~\bibnamefont {Senthil}}, \bibinfo {author} {\bibfnamefont {S.}~\bibnamefont {Sachdev}},\ and\ \bibinfo {author} {\bibfnamefont {M.}~\bibnamefont {Vojta}},\ }\bibfield  {title} {\bibinfo {title} {Fractionalized fermi liquids},\ }\href {https://doi.org/10.1103/PhysRevLett.90.216403} {\bibfield  {journal} {\bibinfo  {journal} {Phys. Rev. Lett.}\ }\textbf {\bibinfo {volume} {90}},\ \bibinfo {pages} {216403} (\bibinfo {year} {2003})}\BibitemShut {NoStop}%
\bibitem [{\citenamefont {Senthil}\ \emph {et~al.}(2004)\citenamefont {Senthil}, \citenamefont {Vojta},\ and\ \citenamefont {Sachdev}}]{Senthil_FL*_PRB}%
  \BibitemOpen
  \bibfield  {author} {\bibinfo {author} {\bibfnamefont {T.}~\bibnamefont {Senthil}}, \bibinfo {author} {\bibfnamefont {M.}~\bibnamefont {Vojta}},\ and\ \bibinfo {author} {\bibfnamefont {S.}~\bibnamefont {Sachdev}},\ }\bibfield  {title} {\bibinfo {title} {Weak magnetism and non-fermi liquids near heavy-fermion critical points},\ }\href {https://doi.org/10.1103/PhysRevB.69.035111} {\bibfield  {journal} {\bibinfo  {journal} {Phys. Rev. B}\ }\textbf {\bibinfo {volume} {69}},\ \bibinfo {pages} {035111} (\bibinfo {year} {2004})}\BibitemShut {NoStop}%
\bibitem [{\citenamefont {Zhao}\ \emph {et~al.}(2023)\citenamefont {Zhao}, \citenamefont {Shen}, \citenamefont {Tao}, \citenamefont {Han}, \citenamefont {Kang}, \citenamefont {Watanabe}, \citenamefont {Taniguchi}, \citenamefont {Mak},\ and\ \citenamefont {Shan}}]{Zhao_Kondo23}%
  \BibitemOpen
  \bibfield  {author} {\bibinfo {author} {\bibfnamefont {W.}~\bibnamefont {Zhao}}, \bibinfo {author} {\bibfnamefont {B.}~\bibnamefont {Shen}}, \bibinfo {author} {\bibfnamefont {Z.}~\bibnamefont {Tao}}, \bibinfo {author} {\bibfnamefont {Z.}~\bibnamefont {Han}}, \bibinfo {author} {\bibfnamefont {K.}~\bibnamefont {Kang}}, \bibinfo {author} {\bibfnamefont {K.}~\bibnamefont {Watanabe}}, \bibinfo {author} {\bibfnamefont {T.}~\bibnamefont {Taniguchi}}, \bibinfo {author} {\bibfnamefont {K.~F.}\ \bibnamefont {Mak}},\ and\ \bibinfo {author} {\bibfnamefont {J.}~\bibnamefont {Shan}},\ }\bibfield  {title} {\bibinfo {title} {Gate-tunable heavy fermions in a moiré kondo lattice},\ }\href {https://doi.org/10.1038/s41586-023-05800-7} {\bibfield  {journal} {\bibinfo  {journal} {Nature}\ }\textbf {\bibinfo {volume} {616}},\ \bibinfo {pages} {61–65} (\bibinfo {year} {2023})}\BibitemShut {NoStop}%
\bibitem [{\citenamefont {Zhao}\ \emph {et~al.}(2024)\citenamefont {Zhao}, \citenamefont {Shen}, \citenamefont {Tao}, \citenamefont {Kim}, \citenamefont {Knüppel}, \citenamefont {Han}, \citenamefont {Zhang}, \citenamefont {Watanabe}, \citenamefont {Taniguchi}, \citenamefont {Chowdhury}, \citenamefont {Shan},\ and\ \citenamefont {Mak}}]{Zhao_2024}%
  \BibitemOpen
  \bibfield  {author} {\bibinfo {author} {\bibfnamefont {W.}~\bibnamefont {Zhao}}, \bibinfo {author} {\bibfnamefont {B.}~\bibnamefont {Shen}}, \bibinfo {author} {\bibfnamefont {Z.}~\bibnamefont {Tao}}, \bibinfo {author} {\bibfnamefont {S.}~\bibnamefont {Kim}}, \bibinfo {author} {\bibfnamefont {P.}~\bibnamefont {Knüppel}}, \bibinfo {author} {\bibfnamefont {Z.}~\bibnamefont {Han}}, \bibinfo {author} {\bibfnamefont {Y.}~\bibnamefont {Zhang}}, \bibinfo {author} {\bibfnamefont {K.}~\bibnamefont {Watanabe}}, \bibinfo {author} {\bibfnamefont {T.}~\bibnamefont {Taniguchi}}, \bibinfo {author} {\bibfnamefont {D.}~\bibnamefont {Chowdhury}}, \bibinfo {author} {\bibfnamefont {J.}~\bibnamefont {Shan}},\ and\ \bibinfo {author} {\bibfnamefont {K.~F.}\ \bibnamefont {Mak}},\ }\bibfield  {title} {\bibinfo {title} {Emergence of ferromagnetism at the onset of moiré kondo breakdown},\ }\href {https://doi.org/10.1038/s41567-024-02636-4} {\bibfield  {journal} {\bibinfo  {journal} {Nature Physics}\ }\textbf {\bibinfo {volume}
  {20}},\ \bibinfo {pages} {1772–1777} (\bibinfo {year} {2024})}\BibitemShut {NoStop}%
\bibitem [{\citenamefont {Zhao}\ \emph {et~al.}(2025)\citenamefont {Zhao}, \citenamefont {Tao}, \citenamefont {Zhang}, \citenamefont {Shen}, \citenamefont {Han}, \citenamefont {Kn{\"u}ppel}, \citenamefont {Zeng}, \citenamefont {Xia}, \citenamefont {Watanabe}, \citenamefont {Taniguchi} \emph {et~al.}}]{zhao_chernmetal}%
  \BibitemOpen
  \bibfield  {author} {\bibinfo {author} {\bibfnamefont {W.}~\bibnamefont {Zhao}}, \bibinfo {author} {\bibfnamefont {Z.}~\bibnamefont {Tao}}, \bibinfo {author} {\bibfnamefont {Y.}~\bibnamefont {Zhang}}, \bibinfo {author} {\bibfnamefont {B.}~\bibnamefont {Shen}}, \bibinfo {author} {\bibfnamefont {Z.}~\bibnamefont {Han}}, \bibinfo {author} {\bibfnamefont {P.}~\bibnamefont {Kn{\"u}ppel}}, \bibinfo {author} {\bibfnamefont {Y.}~\bibnamefont {Zeng}}, \bibinfo {author} {\bibfnamefont {Z.}~\bibnamefont {Xia}}, \bibinfo {author} {\bibfnamefont {K.}~\bibnamefont {Watanabe}}, \bibinfo {author} {\bibfnamefont {T.}~\bibnamefont {Taniguchi}}, \emph {et~al.},\ }\bibfield  {title} {\bibinfo {title} {Emergence of chern metal in a moir$\backslash$'e kondo lattice},\ }\href@noop {} {\bibfield  {journal} {\bibinfo  {journal} {arXiv preprint arXiv:2506.14063}\ } (\bibinfo {year} {2025})}\BibitemShut {NoStop}%
\bibitem [{\citenamefont {Han}\ \emph {et~al.}(2025)\citenamefont {Han}, \citenamefont {Xia}, \citenamefont {Xia}, \citenamefont {Zhao}, \citenamefont {Zhang}, \citenamefont {Watanabe}, \citenamefont {Taniguchi}, \citenamefont {Shan},\ and\ \citenamefont {Mak}}]{Han_TKI}%
  \BibitemOpen
  \bibfield  {author} {\bibinfo {author} {\bibfnamefont {Z.}~\bibnamefont {Han}}, \bibinfo {author} {\bibfnamefont {Y.}~\bibnamefont {Xia}}, \bibinfo {author} {\bibfnamefont {Z.}~\bibnamefont {Xia}}, \bibinfo {author} {\bibfnamefont {W.}~\bibnamefont {Zhao}}, \bibinfo {author} {\bibfnamefont {Y.}~\bibnamefont {Zhang}}, \bibinfo {author} {\bibfnamefont {K.}~\bibnamefont {Watanabe}}, \bibinfo {author} {\bibfnamefont {T.}~\bibnamefont {Taniguchi}}, \bibinfo {author} {\bibfnamefont {J.}~\bibnamefont {Shan}},\ and\ \bibinfo {author} {\bibfnamefont {K.~F.}\ \bibnamefont {Mak}},\ }\bibfield  {title} {\bibinfo {title} {Evidence of topological kondo insulating state in mote2/wse2 moir$\backslash$'e bilayers},\ }\href@noop {} {\bibfield  {journal} {\bibinfo  {journal} {arXiv preprint arXiv:2507.03287}\ } (\bibinfo {year} {2025})}\BibitemShut {NoStop}%
\bibitem [{\citenamefont {Kumar}\ \emph {et~al.}(2022)\citenamefont {Kumar}, \citenamefont {Hu}, \citenamefont {MacDonald},\ and\ \citenamefont {Potter}}]{Kuma_PRB_Kondo}%
  \BibitemOpen
  \bibfield  {author} {\bibinfo {author} {\bibfnamefont {A.}~\bibnamefont {Kumar}}, \bibinfo {author} {\bibfnamefont {N.~C.}\ \bibnamefont {Hu}}, \bibinfo {author} {\bibfnamefont {A.~H.}\ \bibnamefont {MacDonald}},\ and\ \bibinfo {author} {\bibfnamefont {A.~C.}\ \bibnamefont {Potter}},\ }\bibfield  {title} {\bibinfo {title} {Gate-tunable heavy fermion quantum criticality in a moir\'e kondo lattice},\ }\href {https://doi.org/10.1103/PhysRevB.106.L041116} {\bibfield  {journal} {\bibinfo  {journal} {Phys. Rev. B}\ }\textbf {\bibinfo {volume} {106}},\ \bibinfo {pages} {L041116} (\bibinfo {year} {2022})}\BibitemShut {NoStop}%
\bibitem [{\citenamefont {Dalal}\ and\ \citenamefont {Ruhman}(2021)}]{Ruhman_Kondo}%
  \BibitemOpen
  \bibfield  {author} {\bibinfo {author} {\bibfnamefont {A.}~\bibnamefont {Dalal}}\ and\ \bibinfo {author} {\bibfnamefont {J.}~\bibnamefont {Ruhman}},\ }\bibfield  {title} {\bibinfo {title} {Orbitally selective mott phase in electron-doped twisted transition metal-dichalcogenides: A possible realization of the kondo lattice model},\ }\href {https://doi.org/10.1103/PhysRevResearch.3.043173} {\bibfield  {journal} {\bibinfo  {journal} {Phys. Rev. Res.}\ }\textbf {\bibinfo {volume} {3}},\ \bibinfo {pages} {043173} (\bibinfo {year} {2021})}\BibitemShut {NoStop}%
\bibitem [{\citenamefont {Guerci}\ \emph {et~al.}(2023)\citenamefont {Guerci}, \citenamefont {Wang}, \citenamefont {Zang}, \citenamefont {Cano}, \citenamefont {Pixley},\ and\ \citenamefont {Millis}}]{Guerci_2023}%
  \BibitemOpen
  \bibfield  {author} {\bibinfo {author} {\bibfnamefont {D.}~\bibnamefont {Guerci}}, \bibinfo {author} {\bibfnamefont {J.}~\bibnamefont {Wang}}, \bibinfo {author} {\bibfnamefont {J.}~\bibnamefont {Zang}}, \bibinfo {author} {\bibfnamefont {J.}~\bibnamefont {Cano}}, \bibinfo {author} {\bibfnamefont {J.~H.}\ \bibnamefont {Pixley}},\ and\ \bibinfo {author} {\bibfnamefont {A.}~\bibnamefont {Millis}},\ }\bibfield  {title} {\bibinfo {title} {Chiral kondo lattice in doped mote 2 /wse 2 bilayers},\ }\bibfield  {journal} {\bibinfo  {journal} {Science Advances}\ }\textbf {\bibinfo {volume} {9}},\ \href {https://doi.org/10.1126/sciadv.ade7701} {10.1126/sciadv.ade7701} (\bibinfo {year} {2023})\BibitemShut {NoStop}%
\bibitem [{\citenamefont {Xie}\ \emph {et~al.}(2024)\citenamefont {Xie}, \citenamefont {Chen},\ and\ \citenamefont {Si}}]{Qimiao_Kondo}%
  \BibitemOpen
  \bibfield  {author} {\bibinfo {author} {\bibfnamefont {F.}~\bibnamefont {Xie}}, \bibinfo {author} {\bibfnamefont {L.}~\bibnamefont {Chen}},\ and\ \bibinfo {author} {\bibfnamefont {Q.}~\bibnamefont {Si}},\ }\bibfield  {title} {\bibinfo {title} {Kondo effect and its destruction in heterobilayer transition metal dichalcogenides},\ }\href {https://doi.org/10.1103/PhysRevResearch.6.013219} {\bibfield  {journal} {\bibinfo  {journal} {Phys. Rev. Res.}\ }\textbf {\bibinfo {volume} {6}},\ \bibinfo {pages} {013219} (\bibinfo {year} {2024})}\BibitemShut {NoStop}%
\bibitem [{\citenamefont {Mendez-Valderrama}\ \emph {et~al.}(2024)\citenamefont {Mendez-Valderrama}, \citenamefont {Kim},\ and\ \citenamefont {Chowdhury}}]{Kim_TMVI}%
  \BibitemOpen
  \bibfield  {author} {\bibinfo {author} {\bibfnamefont {J.~F.}\ \bibnamefont {Mendez-Valderrama}}, \bibinfo {author} {\bibfnamefont {S.}~\bibnamefont {Kim}},\ and\ \bibinfo {author} {\bibfnamefont {D.}~\bibnamefont {Chowdhury}},\ }\bibfield  {title} {\bibinfo {title} {Correlated topological mixed-valence insulators in moir\'e heterobilayers},\ }\href {https://doi.org/10.1103/PhysRevB.110.L201105} {\bibfield  {journal} {\bibinfo  {journal} {Phys. Rev. B}\ }\textbf {\bibinfo {volume} {110}},\ \bibinfo {pages} {L201105} (\bibinfo {year} {2024})}\BibitemShut {NoStop}%
\bibitem [{\citenamefont {Guerci}\ \emph {et~al.}(2024)\citenamefont {Guerci}, \citenamefont {Lucht}, \citenamefont {Cr\'epel}, \citenamefont {Cano}, \citenamefont {Pixley},\ and\ \citenamefont {Millis}}]{Guerci_TKI}%
  \BibitemOpen
  \bibfield  {author} {\bibinfo {author} {\bibfnamefont {D.}~\bibnamefont {Guerci}}, \bibinfo {author} {\bibfnamefont {K.~P.}\ \bibnamefont {Lucht}}, \bibinfo {author} {\bibfnamefont {V.}~\bibnamefont {Cr\'epel}}, \bibinfo {author} {\bibfnamefont {J.}~\bibnamefont {Cano}}, \bibinfo {author} {\bibfnamefont {J.~H.}\ \bibnamefont {Pixley}},\ and\ \bibinfo {author} {\bibfnamefont {A.}~\bibnamefont {Millis}},\ }\bibfield  {title} {\bibinfo {title} {Topological kondo semimetal and insulator in ab-stacked heterobilayer transition metal dichalcogenides},\ }\href {https://doi.org/10.1103/PhysRevB.110.165128} {\bibfield  {journal} {\bibinfo  {journal} {Phys. Rev. B}\ }\textbf {\bibinfo {volume} {110}},\ \bibinfo {pages} {165128} (\bibinfo {year} {2024})}\BibitemShut {NoStop}%
\bibitem [{\citenamefont {Xie}\ \emph {et~al.}(2025)\citenamefont {Xie}, \citenamefont {Chen}, \citenamefont {Fang},\ and\ \citenamefont {Si}}]{Qimiao_TKS}%
  \BibitemOpen
  \bibfield  {author} {\bibinfo {author} {\bibfnamefont {F.}~\bibnamefont {Xie}}, \bibinfo {author} {\bibfnamefont {L.}~\bibnamefont {Chen}}, \bibinfo {author} {\bibfnamefont {Y.}~\bibnamefont {Fang}},\ and\ \bibinfo {author} {\bibfnamefont {Q.}~\bibnamefont {Si}},\ }\bibfield  {title} {\bibinfo {title} {Topological kondo semimetals emulated in heterobilayer transition metal dichalcogenides},\ }\href {https://doi.org/10.1103/fwbg-kdb9} {\bibfield  {journal} {\bibinfo  {journal} {Phys. Rev. Res.}\ }\textbf {\bibinfo {volume} {7}},\ \bibinfo {pages} {033093} (\bibinfo {year} {2025})}\BibitemShut {NoStop}%
\bibitem [{\citenamefont {Coleman}(1983)}]{Coleman_largeN}%
  \BibitemOpen
  \bibfield  {author} {\bibinfo {author} {\bibfnamefont {P.}~\bibnamefont {Coleman}},\ }\bibfield  {title} {\bibinfo {title} {$\frac{1}{N}$ expansion for the kondo lattice},\ }\href {https://doi.org/10.1103/PhysRevB.28.5255} {\bibfield  {journal} {\bibinfo  {journal} {Phys. Rev. B}\ }\textbf {\bibinfo {volume} {28}},\ \bibinfo {pages} {5255} (\bibinfo {year} {1983})}\BibitemShut {NoStop}%
\bibitem [{\citenamefont {Read}\ \emph {et~al.}(1984)\citenamefont {Read}, \citenamefont {Newns},\ and\ \citenamefont {Doniach}}]{Read_Doniach_largeN}%
  \BibitemOpen
  \bibfield  {author} {\bibinfo {author} {\bibfnamefont {N.}~\bibnamefont {Read}}, \bibinfo {author} {\bibfnamefont {D.~M.}\ \bibnamefont {Newns}},\ and\ \bibinfo {author} {\bibfnamefont {S.}~\bibnamefont {Doniach}},\ }\bibfield  {title} {\bibinfo {title} {Stability of the kondo lattice in the large-$n$ limit},\ }\href {https://doi.org/10.1103/PhysRevB.30.3841} {\bibfield  {journal} {\bibinfo  {journal} {Phys. Rev. B}\ }\textbf {\bibinfo {volume} {30}},\ \bibinfo {pages} {3841} (\bibinfo {year} {1984})}\BibitemShut {NoStop}%
\bibitem [{\citenamefont {Auerbach}\ and\ \citenamefont {Levin}(1986)}]{Auerbach_largeN}%
  \BibitemOpen
  \bibfield  {author} {\bibinfo {author} {\bibfnamefont {A.}~\bibnamefont {Auerbach}}\ and\ \bibinfo {author} {\bibfnamefont {K.}~\bibnamefont {Levin}},\ }\bibfield  {title} {\bibinfo {title} {Kondo bosons and the kondo lattice: Microscopic basis for the heavy fermi liquid},\ }\href {https://doi.org/10.1103/PhysRevLett.57.877} {\bibfield  {journal} {\bibinfo  {journal} {Phys. Rev. Lett.}\ }\textbf {\bibinfo {volume} {57}},\ \bibinfo {pages} {877} (\bibinfo {year} {1986})}\BibitemShut {NoStop}%
\bibitem [{\citenamefont {Millis}\ and\ \citenamefont {Lee}(1987)}]{Millis_largeN}%
  \BibitemOpen
  \bibfield  {author} {\bibinfo {author} {\bibfnamefont {A.~J.}\ \bibnamefont {Millis}}\ and\ \bibinfo {author} {\bibfnamefont {P.~A.}\ \bibnamefont {Lee}},\ }\bibfield  {title} {\bibinfo {title} {Large-orbital-degeneracy expansion for the lattice anderson model},\ }\href {https://doi.org/10.1103/PhysRevB.35.3394} {\bibfield  {journal} {\bibinfo  {journal} {Phys. Rev. B}\ }\textbf {\bibinfo {volume} {35}},\ \bibinfo {pages} {3394} (\bibinfo {year} {1987})}\BibitemShut {NoStop}%
\bibitem [{\citenamefont {Raczkowski}\ and\ \citenamefont {Assaad}(2020)}]{Assaad_SUN_Kondo}%
  \BibitemOpen
  \bibfield  {author} {\bibinfo {author} {\bibfnamefont {M.}~\bibnamefont {Raczkowski}}\ and\ \bibinfo {author} {\bibfnamefont {F.~F.}\ \bibnamefont {Assaad}},\ }\bibfield  {title} {\bibinfo {title} {Phase diagram and dynamics of the $\mathrm{SU}(n)$ symmetric kondo lattice model},\ }\href {https://doi.org/10.1103/PhysRevResearch.2.013276} {\bibfield  {journal} {\bibinfo  {journal} {Phys. Rev. Res.}\ }\textbf {\bibinfo {volume} {2}},\ \bibinfo {pages} {013276} (\bibinfo {year} {2020})}\BibitemShut {NoStop}%
\bibitem [{\citenamefont {Affleck}\ and\ \citenamefont {Marston}(1988)}]{Affleck_largeN_1}%
  \BibitemOpen
  \bibfield  {author} {\bibinfo {author} {\bibfnamefont {I.}~\bibnamefont {Affleck}}\ and\ \bibinfo {author} {\bibfnamefont {J.~B.}\ \bibnamefont {Marston}},\ }\bibfield  {title} {\bibinfo {title} {Large-n limit of the heisenberg-hubbard model: Implications for high-${T}_{c}$ superconductors},\ }\href {https://doi.org/10.1103/PhysRevB.37.3774} {\bibfield  {journal} {\bibinfo  {journal} {Phys. Rev. B}\ }\textbf {\bibinfo {volume} {37}},\ \bibinfo {pages} {3774} (\bibinfo {year} {1988})}\BibitemShut {NoStop}%
\bibitem [{\citenamefont {Marston}\ and\ \citenamefont {Affleck}(1989)}]{Affleck_largeN_2}%
  \BibitemOpen
  \bibfield  {author} {\bibinfo {author} {\bibfnamefont {J.~B.}\ \bibnamefont {Marston}}\ and\ \bibinfo {author} {\bibfnamefont {I.}~\bibnamefont {Affleck}},\ }\bibfield  {title} {\bibinfo {title} {Large-$n$ limit of the hubbard-heisenberg model},\ }\href {https://doi.org/10.1103/PhysRevB.39.11538} {\bibfield  {journal} {\bibinfo  {journal} {Phys. Rev. B}\ }\textbf {\bibinfo {volume} {39}},\ \bibinfo {pages} {11538} (\bibinfo {year} {1989})}\BibitemShut {NoStop}%
\bibitem [{\citenamefont {Read}\ and\ \citenamefont {Sachdev}(1989{\natexlab{a}})}]{Read_Sachdev_1989}%
  \BibitemOpen
  \bibfield  {author} {\bibinfo {author} {\bibfnamefont {N.}~\bibnamefont {Read}}\ and\ \bibinfo {author} {\bibfnamefont {S.}~\bibnamefont {Sachdev}},\ }\bibfield  {title} {\bibinfo {title} {Some features of the phase diagram of the square lattice su(n) antiferromagnet},\ }\href {https://doi.org/10.1016/0550-3213(89)90061-8} {\bibfield  {journal} {\bibinfo  {journal} {Nuclear Physics B}\ }\textbf {\bibinfo {volume} {316}},\ \bibinfo {pages} {609–640} (\bibinfo {year} {1989}{\natexlab{a}})}\BibitemShut {NoStop}%
\bibitem [{\citenamefont {Read}\ and\ \citenamefont {Sachdev}(1989{\natexlab{b}})}]{Read_Sachdev_PRL1989}%
  \BibitemOpen
  \bibfield  {author} {\bibinfo {author} {\bibfnamefont {N.}~\bibnamefont {Read}}\ and\ \bibinfo {author} {\bibfnamefont {S.}~\bibnamefont {Sachdev}},\ }\bibfield  {title} {\bibinfo {title} {Valence-bond and spin-peierls ground states of low-dimensional quantum antiferromagnets},\ }\href {https://doi.org/10.1103/PhysRevLett.62.1694} {\bibfield  {journal} {\bibinfo  {journal} {Phys. Rev. Lett.}\ }\textbf {\bibinfo {volume} {62}},\ \bibinfo {pages} {1694} (\bibinfo {year} {1989}{\natexlab{b}})}\BibitemShut {NoStop}%
\bibitem [{\citenamefont {Read}\ and\ \citenamefont {Sachdev}(1990)}]{Read_Sachdev_PRB1990}%
  \BibitemOpen
  \bibfield  {author} {\bibinfo {author} {\bibfnamefont {N.}~\bibnamefont {Read}}\ and\ \bibinfo {author} {\bibfnamefont {S.}~\bibnamefont {Sachdev}},\ }\bibfield  {title} {\bibinfo {title} {Spin-peierls, valence-bond solid, and n\'eel ground states of low-dimensional quantum antiferromagnets},\ }\href {https://doi.org/10.1103/PhysRevB.42.4568} {\bibfield  {journal} {\bibinfo  {journal} {Phys. Rev. B}\ }\textbf {\bibinfo {volume} {42}},\ \bibinfo {pages} {4568} (\bibinfo {year} {1990})}\BibitemShut {NoStop}%
\bibitem [{\citenamefont {Read}\ and\ \citenamefont {Sachdev}(1991)}]{Read_Sachdev_PRL1991}%
  \BibitemOpen
  \bibfield  {author} {\bibinfo {author} {\bibfnamefont {N.}~\bibnamefont {Read}}\ and\ \bibinfo {author} {\bibfnamefont {S.}~\bibnamefont {Sachdev}},\ }\bibfield  {title} {\bibinfo {title} {Large-n expansion for frustrated quantum antiferromagnets},\ }\href {https://doi.org/10.1103/PhysRevLett.66.1773} {\bibfield  {journal} {\bibinfo  {journal} {Phys. Rev. Lett.}\ }\textbf {\bibinfo {volume} {66}},\ \bibinfo {pages} {1773} (\bibinfo {year} {1991})}\BibitemShut {NoStop}%
\bibitem [{\citenamefont {Sachdev}(1992)}]{Sachdev_PRB1992}%
  \BibitemOpen
  \bibfield  {author} {\bibinfo {author} {\bibfnamefont {S.}~\bibnamefont {Sachdev}},\ }\bibfield  {title} {\bibinfo {title} {Kagome\ifmmode\acute\else\textasciiacute\fi{}- and triangular-lattice heisenberg antiferromagnets: Ordering from quantum fluctuations and quantum-disordered ground states with unconfined bosonic spinons},\ }\href {https://doi.org/10.1103/PhysRevB.45.12377} {\bibfield  {journal} {\bibinfo  {journal} {Phys. Rev. B}\ }\textbf {\bibinfo {volume} {45}},\ \bibinfo {pages} {12377} (\bibinfo {year} {1992})}\BibitemShut {NoStop}%
\bibitem [{\citenamefont {Wu}\ \emph {et~al.}(2003)\citenamefont {Wu}, \citenamefont {Hu},\ and\ \citenamefont {Zhang}}]{SCZhang_SO5}%
  \BibitemOpen
  \bibfield  {author} {\bibinfo {author} {\bibfnamefont {C.}~\bibnamefont {Wu}}, \bibinfo {author} {\bibfnamefont {J.-p.}\ \bibnamefont {Hu}},\ and\ \bibinfo {author} {\bibfnamefont {S.-c.}\ \bibnamefont {Zhang}},\ }\bibfield  {title} {\bibinfo {title} {Exact so(5) symmetry in the spin-$3/2$ fermionic system},\ }\href {https://doi.org/10.1103/PhysRevLett.91.186402} {\bibfield  {journal} {\bibinfo  {journal} {Phys. Rev. Lett.}\ }\textbf {\bibinfo {volume} {91}},\ \bibinfo {pages} {186402} (\bibinfo {year} {2003})}\BibitemShut {NoStop}%
\bibitem [{\citenamefont {Assaad}(2005)}]{Assaad_SUN_QMC}%
  \BibitemOpen
  \bibfield  {author} {\bibinfo {author} {\bibfnamefont {F.~F.}\ \bibnamefont {Assaad}},\ }\bibfield  {title} {\bibinfo {title} {Phase diagram of the half-filled two-dimensional $\mathrm{SU}(n)$ hubbard-heisenberg model: A quantum monte carlo study},\ }\href {https://doi.org/10.1103/PhysRevB.71.075103} {\bibfield  {journal} {\bibinfo  {journal} {Phys. Rev. B}\ }\textbf {\bibinfo {volume} {71}},\ \bibinfo {pages} {075103} (\bibinfo {year} {2005})}\BibitemShut {NoStop}%
\bibitem [{\citenamefont {Arovas}(2008)}]{Arovas_SUNsinglet}%
  \BibitemOpen
  \bibfield  {author} {\bibinfo {author} {\bibfnamefont {D.~P.}\ \bibnamefont {Arovas}},\ }\bibfield  {title} {\bibinfo {title} {Simplex solid states of $\mathrm{SU}(n)$ quantum antiferromagnets},\ }\href {https://doi.org/10.1103/PhysRevB.77.104404} {\bibfield  {journal} {\bibinfo  {journal} {Phys. Rev. B}\ }\textbf {\bibinfo {volume} {77}},\ \bibinfo {pages} {104404} (\bibinfo {year} {2008})}\BibitemShut {NoStop}%
\bibitem [{\citenamefont {Xu}\ and\ \citenamefont {Wu}(2008)}]{Xu_SU4Plaquette}%
  \BibitemOpen
  \bibfield  {author} {\bibinfo {author} {\bibfnamefont {C.}~\bibnamefont {Xu}}\ and\ \bibinfo {author} {\bibfnamefont {C.}~\bibnamefont {Wu}},\ }\bibfield  {title} {\bibinfo {title} {Resonating plaquette phases in su(4) heisenberg antiferromagnet},\ }\href {https://doi.org/10.1103/PhysRevB.77.134449} {\bibfield  {journal} {\bibinfo  {journal} {Phys. Rev. B}\ }\textbf {\bibinfo {volume} {77}},\ \bibinfo {pages} {134449} (\bibinfo {year} {2008})}\BibitemShut {NoStop}%
\bibitem [{\citenamefont {Hermele}\ \emph {et~al.}(2009)\citenamefont {Hermele}, \citenamefont {Gurarie},\ and\ \citenamefont {Rey}}]{Hermele_SU4plaquette}%
  \BibitemOpen
  \bibfield  {author} {\bibinfo {author} {\bibfnamefont {M.}~\bibnamefont {Hermele}}, \bibinfo {author} {\bibfnamefont {V.}~\bibnamefont {Gurarie}},\ and\ \bibinfo {author} {\bibfnamefont {A.~M.}\ \bibnamefont {Rey}},\ }\bibfield  {title} {\bibinfo {title} {Mott insulators of ultracold fermionic alkaline earth atoms: Underconstrained magnetism and chiral spin liquid},\ }\href {https://doi.org/10.1103/PhysRevLett.103.135301} {\bibfield  {journal} {\bibinfo  {journal} {Phys. Rev. Lett.}\ }\textbf {\bibinfo {volume} {103}},\ \bibinfo {pages} {135301} (\bibinfo {year} {2009})}\BibitemShut {NoStop}%
\bibitem [{\citenamefont {Gorshkov}\ \emph {et~al.}(2010)\citenamefont {Gorshkov}, \citenamefont {Hermele}, \citenamefont {Gurarie}, \citenamefont {Xu}, \citenamefont {Julienne}, \citenamefont {Ye}, \citenamefont {Zoller}, \citenamefont {Demler}, \citenamefont {Lukin},\ and\ \citenamefont {Rey}}]{Gorshkov_Cold_SUN}%
  \BibitemOpen
  \bibfield  {author} {\bibinfo {author} {\bibfnamefont {A.~V.}\ \bibnamefont {Gorshkov}}, \bibinfo {author} {\bibfnamefont {M.}~\bibnamefont {Hermele}}, \bibinfo {author} {\bibfnamefont {V.}~\bibnamefont {Gurarie}}, \bibinfo {author} {\bibfnamefont {C.}~\bibnamefont {Xu}}, \bibinfo {author} {\bibfnamefont {P.~S.}\ \bibnamefont {Julienne}}, \bibinfo {author} {\bibfnamefont {J.}~\bibnamefont {Ye}}, \bibinfo {author} {\bibfnamefont {P.}~\bibnamefont {Zoller}}, \bibinfo {author} {\bibfnamefont {E.}~\bibnamefont {Demler}}, \bibinfo {author} {\bibfnamefont {M.~D.}\ \bibnamefont {Lukin}},\ and\ \bibinfo {author} {\bibfnamefont {A.~M.}\ \bibnamefont {Rey}},\ }\bibfield  {title} {\bibinfo {title} {Two-orbital s u(n) magnetism with ultracold alkaline-earth atoms},\ }\href {https://doi.org/10.1038/nphys1535} {\bibfield  {journal} {\bibinfo  {journal} {Nature Physics}\ }\textbf {\bibinfo {volume} {6}},\ \bibinfo {pages} {289–295} (\bibinfo {year} {2010})}\BibitemShut {NoStop}%
\bibitem [{\citenamefont {Lang}\ \emph {et~al.}(2013)\citenamefont {Lang}, \citenamefont {Meng}, \citenamefont {Muramatsu}, \citenamefont {Wessel},\ and\ \citenamefont {Assaad}}]{Assaad_SUN_plaquette}%
  \BibitemOpen
  \bibfield  {author} {\bibinfo {author} {\bibfnamefont {T.~C.}\ \bibnamefont {Lang}}, \bibinfo {author} {\bibfnamefont {Z.~Y.}\ \bibnamefont {Meng}}, \bibinfo {author} {\bibfnamefont {A.}~\bibnamefont {Muramatsu}}, \bibinfo {author} {\bibfnamefont {S.}~\bibnamefont {Wessel}},\ and\ \bibinfo {author} {\bibfnamefont {F.~F.}\ \bibnamefont {Assaad}},\ }\bibfield  {title} {\bibinfo {title} {Dimerized solids and resonating plaquette order in $\mathrm{SU}(n)$-dirac fermions},\ }\href {https://doi.org/10.1103/PhysRevLett.111.066401} {\bibfield  {journal} {\bibinfo  {journal} {Phys. Rev. Lett.}\ }\textbf {\bibinfo {volume} {111}},\ \bibinfo {pages} {066401} (\bibinfo {year} {2013})}\BibitemShut {NoStop}%
\bibitem [{\citenamefont {Keselman}\ \emph {et~al.}(2020)\citenamefont {Keselman}, \citenamefont {Bauer}, \citenamefont {Xu},\ and\ \citenamefont {Jian}}]{Keselman_su4_prl}%
  \BibitemOpen
  \bibfield  {author} {\bibinfo {author} {\bibfnamefont {A.}~\bibnamefont {Keselman}}, \bibinfo {author} {\bibfnamefont {B.}~\bibnamefont {Bauer}}, \bibinfo {author} {\bibfnamefont {C.}~\bibnamefont {Xu}},\ and\ \bibinfo {author} {\bibfnamefont {C.-M.}\ \bibnamefont {Jian}},\ }\bibfield  {title} {\bibinfo {title} {Emergent fermi surface in a triangular-lattice su(4) quantum antiferromagnet},\ }\href {https://doi.org/10.1103/PhysRevLett.125.117202} {\bibfield  {journal} {\bibinfo  {journal} {Phys. Rev. Lett.}\ }\textbf {\bibinfo {volume} {125}},\ \bibinfo {pages} {117202} (\bibinfo {year} {2020})}\BibitemShut {NoStop}%
\bibitem [{\citenamefont {Bieri}\ \emph {et~al.}(2012)\citenamefont {Bieri}, \citenamefont {Serbyn}, \citenamefont {Senthil},\ and\ \citenamefont {Lee}}]{SU3_pairedCSL}%
  \BibitemOpen
  \bibfield  {author} {\bibinfo {author} {\bibfnamefont {S.}~\bibnamefont {Bieri}}, \bibinfo {author} {\bibfnamefont {M.}~\bibnamefont {Serbyn}}, \bibinfo {author} {\bibfnamefont {T.}~\bibnamefont {Senthil}},\ and\ \bibinfo {author} {\bibfnamefont {P.~A.}\ \bibnamefont {Lee}},\ }\bibfield  {title} {\bibinfo {title} {Paired chiral spin liquid with a fermi surface in $s=1$ model on the triangular lattice},\ }\href {https://doi.org/10.1103/PhysRevB.86.224409} {\bibfield  {journal} {\bibinfo  {journal} {Phys. Rev. B}\ }\textbf {\bibinfo {volume} {86}},\ \bibinfo {pages} {224409} (\bibinfo {year} {2012})}\BibitemShut {NoStop}%
\bibitem [{\citenamefont {Ramires}\ and\ \citenamefont {Lado}(2021)}]{Lado_SU4TTG}%
  \BibitemOpen
  \bibfield  {author} {\bibinfo {author} {\bibfnamefont {A.}~\bibnamefont {Ramires}}\ and\ \bibinfo {author} {\bibfnamefont {J.~L.}\ \bibnamefont {Lado}},\ }\bibfield  {title} {\bibinfo {title} {Emulating heavy fermions in twisted trilayer graphene},\ }\href {https://doi.org/10.1103/PhysRevLett.127.026401} {\bibfield  {journal} {\bibinfo  {journal} {Phys. Rev. Lett.}\ }\textbf {\bibinfo {volume} {127}},\ \bibinfo {pages} {026401} (\bibinfo {year} {2021})}\BibitemShut {NoStop}%
\bibitem [{\citenamefont {Chou}\ and\ \citenamefont {Das~Sarma}(2023)}]{SDS_SU8TBG}%
  \BibitemOpen
  \bibfield  {author} {\bibinfo {author} {\bibfnamefont {Y.-Z.}\ \bibnamefont {Chou}}\ and\ \bibinfo {author} {\bibfnamefont {S.}~\bibnamefont {Das~Sarma}},\ }\bibfield  {title} {\bibinfo {title} {Kondo lattice model in magic-angle twisted bilayer graphene},\ }\href {https://doi.org/10.1103/PhysRevLett.131.026501} {\bibfield  {journal} {\bibinfo  {journal} {Phys. Rev. Lett.}\ }\textbf {\bibinfo {volume} {131}},\ \bibinfo {pages} {026501} (\bibinfo {year} {2023})}\BibitemShut {NoStop}%
\bibitem [{\citenamefont {Song}\ and\ \citenamefont {Bernevig}(2022)}]{Song_Bernevig_THF}%
  \BibitemOpen
  \bibfield  {author} {\bibinfo {author} {\bibfnamefont {Z.-D.}\ \bibnamefont {Song}}\ and\ \bibinfo {author} {\bibfnamefont {B.~A.}\ \bibnamefont {Bernevig}},\ }\bibfield  {title} {\bibinfo {title} {Magic-angle twisted bilayer graphene as a topological heavy fermion problem},\ }\href {https://doi.org/10.1103/PhysRevLett.129.047601} {\bibfield  {journal} {\bibinfo  {journal} {Phys. Rev. Lett.}\ }\textbf {\bibinfo {volume} {129}},\ \bibinfo {pages} {047601} (\bibinfo {year} {2022})}\BibitemShut {NoStop}%
\bibitem [{\citenamefont {Zhang}\ and\ \citenamefont {Vishwanath}(2020)}]{zhang2020electrical}%
  \BibitemOpen
  \bibfield  {author} {\bibinfo {author} {\bibfnamefont {Y.-H.}\ \bibnamefont {Zhang}}\ and\ \bibinfo {author} {\bibfnamefont {A.}~\bibnamefont {Vishwanath}},\ }\bibfield  {title} {\bibinfo {title} {Electrical detection of spin liquids in double moir$\backslash$'e layers},\ }\href@noop {} {\bibfield  {journal} {\bibinfo  {journal} {arXiv preprint arXiv:2005.12925}\ } (\bibinfo {year} {2020})}\BibitemShut {NoStop}%
\bibitem [{\citenamefont {Zhang}\ \emph {et~al.}(2021{\natexlab{a}})\citenamefont {Zhang}, \citenamefont {Sheng},\ and\ \citenamefont {Vishwanath}}]{zhang_prl}%
  \BibitemOpen
  \bibfield  {author} {\bibinfo {author} {\bibfnamefont {Y.-H.}\ \bibnamefont {Zhang}}, \bibinfo {author} {\bibfnamefont {D.~N.}\ \bibnamefont {Sheng}},\ and\ \bibinfo {author} {\bibfnamefont {A.}~\bibnamefont {Vishwanath}},\ }\bibfield  {title} {\bibinfo {title} {Su(4) chiral spin liquid, exciton supersolid, and electric detection in moir\'e bilayers},\ }\href {https://doi.org/10.1103/PhysRevLett.127.247701} {\bibfield  {journal} {\bibinfo  {journal} {Phys. Rev. Lett.}\ }\textbf {\bibinfo {volume} {127}},\ \bibinfo {pages} {247701} (\bibinfo {year} {2021}{\natexlab{a}})}\BibitemShut {NoStop}%
\bibitem [{\citenamefont {Kuhlenkamp}\ \emph {et~al.}(2024)\citenamefont {Kuhlenkamp}, \citenamefont {Kadow}, \citenamefont {Imamo\ifmmode~\breve{g}\else \u{g}\fi{}lu},\ and\ \citenamefont {Knap}}]{Wilhelm_PRX_PSL}%
  \BibitemOpen
  \bibfield  {author} {\bibinfo {author} {\bibfnamefont {C.}~\bibnamefont {Kuhlenkamp}}, \bibinfo {author} {\bibfnamefont {W.}~\bibnamefont {Kadow}}, \bibinfo {author} {\bibfnamefont {A.~m.~c.}\ \bibnamefont {Imamo\ifmmode~\breve{g}\else \u{g}\fi{}lu}},\ and\ \bibinfo {author} {\bibfnamefont {M.}~\bibnamefont {Knap}},\ }\bibfield  {title} {\bibinfo {title} {Chiral pseudospin liquids in moir\'e heterostructures},\ }\href {https://doi.org/10.1103/PhysRevX.14.021013} {\bibfield  {journal} {\bibinfo  {journal} {Phys. Rev. X}\ }\textbf {\bibinfo {volume} {14}},\ \bibinfo {pages} {021013} (\bibinfo {year} {2024})}\BibitemShut {NoStop}%
\bibitem [{\citenamefont {Zhang}\ \emph {et~al.}(2023)\citenamefont {Zhang}, \citenamefont {Zhou},\ and\ \citenamefont {Zhang}}]{Boran_Yahui_SU4}%
  \BibitemOpen
  \bibfield  {author} {\bibinfo {author} {\bibfnamefont {S.}~\bibnamefont {Zhang}}, \bibinfo {author} {\bibfnamefont {B.}~\bibnamefont {Zhou}},\ and\ \bibinfo {author} {\bibfnamefont {Y.-H.}\ \bibnamefont {Zhang}},\ }\bibfield  {title} {\bibinfo {title} {Approximate su (4) spin models on triangular and honeycomb lattices in twisted ab-stacked wse $ \_2 $ homo-bilayer},\ }\href@noop {} {\bibfield  {journal} {\bibinfo  {journal} {arXiv preprint arXiv:2302.07750}\ } (\bibinfo {year} {2023})}\BibitemShut {NoStop}%
\bibitem [{\citenamefont {Del~Re}\ and\ \citenamefont {Classen}(2024)}]{Lorenzo_PRR}%
  \BibitemOpen
  \bibfield  {author} {\bibinfo {author} {\bibfnamefont {L.}~\bibnamefont {Del~Re}}\ and\ \bibinfo {author} {\bibfnamefont {L.}~\bibnamefont {Classen}},\ }\bibfield  {title} {\bibinfo {title} {Field control of symmetry-broken and quantum disordered phases in frustrated moir\'e bilayers with population imbalance},\ }\href {https://doi.org/10.1103/PhysRevResearch.6.023082} {\bibfield  {journal} {\bibinfo  {journal} {Phys. Rev. Res.}\ }\textbf {\bibinfo {volume} {6}},\ \bibinfo {pages} {023082} (\bibinfo {year} {2024})}\BibitemShut {NoStop}%
\bibitem [{\citenamefont {Xu}\ \emph {et~al.}(2022)\citenamefont {Xu}, \citenamefont {Kang}, \citenamefont {Watanabe}, \citenamefont {Taniguchi}, \citenamefont {Mak},\ and\ \citenamefont {Shan}}]{SU4_tWSe2_Expt}%
  \BibitemOpen
  \bibfield  {author} {\bibinfo {author} {\bibfnamefont {Y.}~\bibnamefont {Xu}}, \bibinfo {author} {\bibfnamefont {K.}~\bibnamefont {Kang}}, \bibinfo {author} {\bibfnamefont {K.}~\bibnamefont {Watanabe}}, \bibinfo {author} {\bibfnamefont {T.}~\bibnamefont {Taniguchi}}, \bibinfo {author} {\bibfnamefont {K.~F.}\ \bibnamefont {Mak}},\ and\ \bibinfo {author} {\bibfnamefont {J.}~\bibnamefont {Shan}},\ }\bibfield  {title} {\bibinfo {title} {A tunable bilayer hubbard model in twisted wse2},\ }\href {https://doi.org/10.1038/s41565-022-01180-7} {\bibfield  {journal} {\bibinfo  {journal} {Nature Nanotechnology}\ }\textbf {\bibinfo {volume} {17}},\ \bibinfo {pages} {934–939} (\bibinfo {year} {2022})}\BibitemShut {NoStop}%
\bibitem [{\citenamefont {Borda}\ \emph {et~al.}(2003)\citenamefont {Borda}, \citenamefont {Zar\'and}, \citenamefont {Hofstetter}, \citenamefont {Halperin},\ and\ \citenamefont {von Delft}}]{Borda_DoubleQD}%
  \BibitemOpen
  \bibfield  {author} {\bibinfo {author} {\bibfnamefont {L.}~\bibnamefont {Borda}}, \bibinfo {author} {\bibfnamefont {G.}~\bibnamefont {Zar\'and}}, \bibinfo {author} {\bibfnamefont {W.}~\bibnamefont {Hofstetter}}, \bibinfo {author} {\bibfnamefont {B.~I.}\ \bibnamefont {Halperin}},\ and\ \bibinfo {author} {\bibfnamefont {J.}~\bibnamefont {von Delft}},\ }\bibfield  {title} {\bibinfo {title} {Su(4) fermi liquid state and spin filtering in a double quantum dot system},\ }\href {https://doi.org/10.1103/PhysRevLett.90.026602} {\bibfield  {journal} {\bibinfo  {journal} {Phys. Rev. Lett.}\ }\textbf {\bibinfo {volume} {90}},\ \bibinfo {pages} {026602} (\bibinfo {year} {2003})}\BibitemShut {NoStop}%
\bibitem [{\citenamefont {Kim}\ \emph {et~al.}(2023)\citenamefont {Kim}, \citenamefont {Dominguez}, \citenamefont {Mayorga-Luna}, \citenamefont {Ye}, \citenamefont {Embley}, \citenamefont {Tan}, \citenamefont {Ni}, \citenamefont {Liu}, \citenamefont {Ford}, \citenamefont {Gao}, \citenamefont {Arash}, \citenamefont {Watanabe}, \citenamefont {Taniguchi}, \citenamefont {Kim}, \citenamefont {Shih}, \citenamefont {Lai}, \citenamefont {Yao}, \citenamefont {Yang}, \citenamefont {Li},\ and\ \citenamefont {Miyahara}}]{Kim_thBN_2023}%
  \BibitemOpen
  \bibfield  {author} {\bibinfo {author} {\bibfnamefont {D.~S.}\ \bibnamefont {Kim}}, \bibinfo {author} {\bibfnamefont {R.~C.}\ \bibnamefont {Dominguez}}, \bibinfo {author} {\bibfnamefont {R.}~\bibnamefont {Mayorga-Luna}}, \bibinfo {author} {\bibfnamefont {D.}~\bibnamefont {Ye}}, \bibinfo {author} {\bibfnamefont {J.}~\bibnamefont {Embley}}, \bibinfo {author} {\bibfnamefont {T.}~\bibnamefont {Tan}}, \bibinfo {author} {\bibfnamefont {Y.}~\bibnamefont {Ni}}, \bibinfo {author} {\bibfnamefont {Z.}~\bibnamefont {Liu}}, \bibinfo {author} {\bibfnamefont {M.}~\bibnamefont {Ford}}, \bibinfo {author} {\bibfnamefont {F.~Y.}\ \bibnamefont {Gao}}, \bibinfo {author} {\bibfnamefont {S.}~\bibnamefont {Arash}}, \bibinfo {author} {\bibfnamefont {K.}~\bibnamefont {Watanabe}}, \bibinfo {author} {\bibfnamefont {T.}~\bibnamefont {Taniguchi}}, \bibinfo {author} {\bibfnamefont {S.}~\bibnamefont {Kim}}, \bibinfo {author} {\bibfnamefont {C.-K.}\ \bibnamefont {Shih}}, \bibinfo {author} {\bibfnamefont {K.}~\bibnamefont {Lai}}, \bibinfo
  {author} {\bibfnamefont {W.}~\bibnamefont {Yao}}, \bibinfo {author} {\bibfnamefont {L.}~\bibnamefont {Yang}}, \bibinfo {author} {\bibfnamefont {X.}~\bibnamefont {Li}},\ and\ \bibinfo {author} {\bibfnamefont {Y.}~\bibnamefont {Miyahara}},\ }\bibfield  {title} {\bibinfo {title} {Electrostatic moiré potential from twisted hexagonal boron nitride layers},\ }\href {https://doi.org/10.1038/s41563-023-01637-7} {\bibfield  {journal} {\bibinfo  {journal} {Nature Materials}\ }\textbf {\bibinfo {volume} {23}},\ \bibinfo {pages} {65–70} (\bibinfo {year} {2023})}\BibitemShut {NoStop}%
\bibitem [{\citenamefont {Schrieffer}\ and\ \citenamefont {Wolff}(1966)}]{SW_prl}%
  \BibitemOpen
  \bibfield  {author} {\bibinfo {author} {\bibfnamefont {J.~R.}\ \bibnamefont {Schrieffer}}\ and\ \bibinfo {author} {\bibfnamefont {P.~A.}\ \bibnamefont {Wolff}},\ }\bibfield  {title} {\bibinfo {title} {Relation between the anderson and kondo hamiltonians},\ }\href {https://doi.org/10.1103/PhysRev.149.491} {\bibfield  {journal} {\bibinfo  {journal} {Phys. Rev.}\ }\textbf {\bibinfo {volume} {149}},\ \bibinfo {pages} {491} (\bibinfo {year} {1966})}\BibitemShut {NoStop}%
\bibitem [{\citenamefont {MacDonald}\ \emph {et~al.}(1988)\citenamefont {MacDonald}, \citenamefont {Girvin},\ and\ \citenamefont {Yoshioka}}]{MacDonald_tU}%
  \BibitemOpen
  \bibfield  {author} {\bibinfo {author} {\bibfnamefont {A.~H.}\ \bibnamefont {MacDonald}}, \bibinfo {author} {\bibfnamefont {S.~M.}\ \bibnamefont {Girvin}},\ and\ \bibinfo {author} {\bibfnamefont {D.}~\bibnamefont {Yoshioka}},\ }\bibfield  {title} {\bibinfo {title} {$\frac{t}{U}$ expansion for the hubbard model},\ }\href {https://doi.org/10.1103/PhysRevB.37.9753} {\bibfield  {journal} {\bibinfo  {journal} {Phys. Rev. B}\ }\textbf {\bibinfo {volume} {37}},\ \bibinfo {pages} {9753} (\bibinfo {year} {1988})}\BibitemShut {NoStop}%
\bibitem [{si()}]{si}%
  \BibitemOpen
  \href@noop {} {}\bibinfo {note} {See supplementary material for additional details on the Schrieffer-Wolff transformation, various spontaneous symmetry breakings in the Kondo-unscreened phases.}\BibitemShut {Stop}%
\bibitem [{\citenamefont {Motrunich}(2006)}]{Motrunich_PRB}%
  \BibitemOpen
  \bibfield  {author} {\bibinfo {author} {\bibfnamefont {O.~I.}\ \bibnamefont {Motrunich}},\ }\bibfield  {title} {\bibinfo {title} {Orbital magnetic field effects in spin liquid with spinon fermi sea: Possible application to $\ensuremath{\kappa}\text{\ensuremath{-}}{(\mathrm{ET})}_{2}{\mathrm{cu}}_{2}{(\mathrm{C}\mathrm{N})}_{3}$},\ }\href {https://doi.org/10.1103/PhysRevB.73.155115} {\bibfield  {journal} {\bibinfo  {journal} {Phys. Rev. B}\ }\textbf {\bibinfo {volume} {73}},\ \bibinfo {pages} {155115} (\bibinfo {year} {2006})}\BibitemShut {NoStop}%
\bibitem [{\citenamefont {Kalmeyer}\ and\ \citenamefont {Laughlin}(1987)}]{Laughlin_CSL}%
  \BibitemOpen
  \bibfield  {author} {\bibinfo {author} {\bibfnamefont {V.}~\bibnamefont {Kalmeyer}}\ and\ \bibinfo {author} {\bibfnamefont {R.~B.}\ \bibnamefont {Laughlin}},\ }\bibfield  {title} {\bibinfo {title} {Equivalence of the resonating-valence-bond and fractional quantum hall states},\ }\href {https://doi.org/10.1103/PhysRevLett.59.2095} {\bibfield  {journal} {\bibinfo  {journal} {Phys. Rev. Lett.}\ }\textbf {\bibinfo {volume} {59}},\ \bibinfo {pages} {2095} (\bibinfo {year} {1987})}\BibitemShut {NoStop}%
\bibitem [{\citenamefont {Wen}\ \emph {et~al.}(1989)\citenamefont {Wen}, \citenamefont {Wilczek},\ and\ \citenamefont {Zee}}]{Wen_CSL}%
  \BibitemOpen
  \bibfield  {author} {\bibinfo {author} {\bibfnamefont {X.~G.}\ \bibnamefont {Wen}}, \bibinfo {author} {\bibfnamefont {F.}~\bibnamefont {Wilczek}},\ and\ \bibinfo {author} {\bibfnamefont {A.}~\bibnamefont {Zee}},\ }\bibfield  {title} {\bibinfo {title} {Chiral spin states and superconductivity},\ }\href {https://doi.org/10.1103/PhysRevB.39.11413} {\bibfield  {journal} {\bibinfo  {journal} {Phys. Rev. B}\ }\textbf {\bibinfo {volume} {39}},\ \bibinfo {pages} {11413} (\bibinfo {year} {1989})}\BibitemShut {NoStop}%
\bibitem [{\citenamefont {{Zhang}}\ \emph {et~al.}(2025)\citenamefont {{Zhang}}, \citenamefont {{Zhang}},\ and\ \citenamefont {{Song}}}]{Zhang_Song_SU4SC}%
  \BibitemOpen
  \bibfield  {author} {\bibinfo {author} {\bibfnamefont {L.}~\bibnamefont {{Zhang}}}, \bibinfo {author} {\bibfnamefont {Y.-H.}\ \bibnamefont {{Zhang}}},\ and\ \bibinfo {author} {\bibfnamefont {X.-Y.}\ \bibnamefont {{Song}}},\ }\bibfield  {title} {\bibinfo {title} {{Charge-4$e$ Anyon Superconductor from Doping $\text{SU}(4)_1$ chiral spin liquid}},\ }\href {https://doi.org/10.48550/arXiv.2508.12370} {\bibfield  {journal} {\bibinfo  {journal} {arXiv e-prints}\ ,\ \bibinfo {eid} {arXiv:2508.12370}} (\bibinfo {year} {2025})},\ \Eprint {https://arxiv.org/abs/2508.12370} {arXiv:2508.12370 [cond-mat.str-el]} \BibitemShut {NoStop}%
\bibitem [{\citenamefont {Gresista}\ \emph {et~al.}(2023)\citenamefont {Gresista}, \citenamefont {Kiese}, \citenamefont {Trebst},\ and\ \citenamefont {Scherer}}]{Trebst_brokenSU4}%
  \BibitemOpen
  \bibfield  {author} {\bibinfo {author} {\bibfnamefont {L.}~\bibnamefont {Gresista}}, \bibinfo {author} {\bibfnamefont {D.}~\bibnamefont {Kiese}}, \bibinfo {author} {\bibfnamefont {S.}~\bibnamefont {Trebst}},\ and\ \bibinfo {author} {\bibfnamefont {M.~M.}\ \bibnamefont {Scherer}},\ }\bibfield  {title} {\bibinfo {title} {Spin-valley magnetism on the triangular moir\'e lattice with su(4) breaking interactions},\ }\href {https://doi.org/10.1103/PhysRevB.108.045102} {\bibfield  {journal} {\bibinfo  {journal} {Phys. Rev. B}\ }\textbf {\bibinfo {volume} {108}},\ \bibinfo {pages} {045102} (\bibinfo {year} {2023})}\BibitemShut {NoStop}%
\bibitem [{\citenamefont {{Wu}}\ \emph {et~al.}(2025)\citenamefont {{Wu}}, \citenamefont {{Cui}}, \citenamefont {{Wu}}, \citenamefont {{Meng}}, \citenamefont {{Chen}}, \citenamefont {{Yan}}, \citenamefont {{Ma}}, \citenamefont {{Taniguchi}}, \citenamefont {{Watanabe}}, \citenamefont {{Lin}}, \citenamefont {{Shi}},\ and\ \citenamefont {{Cui}}}]{ShiZeng_KI}%
  \BibitemOpen
  \bibfield  {author} {\bibinfo {author} {\bibfnamefont {Q.}~\bibnamefont {{Wu}}}, \bibinfo {author} {\bibfnamefont {J.}~\bibnamefont {{Cui}}}, \bibinfo {author} {\bibfnamefont {A.-K.}\ \bibnamefont {{Wu}}}, \bibinfo {author} {\bibfnamefont {Y.}~\bibnamefont {{Meng}}}, \bibinfo {author} {\bibfnamefont {D.}~\bibnamefont {{Chen}}}, \bibinfo {author} {\bibfnamefont {L.}~\bibnamefont {{Yan}}}, \bibinfo {author} {\bibfnamefont {L.}~\bibnamefont {{Ma}}}, \bibinfo {author} {\bibfnamefont {T.}~\bibnamefont {{Taniguchi}}}, \bibinfo {author} {\bibfnamefont {K.}~\bibnamefont {{Watanabe}}}, \bibinfo {author} {\bibfnamefont {S.-Z.}\ \bibnamefont {{Lin}}}, \bibinfo {author} {\bibfnamefont {S.-F.}\ \bibnamefont {{Shi}}},\ and\ \bibinfo {author} {\bibfnamefont {Y.-T.}\ \bibnamefont {{Cui}}},\ }\bibfield  {title} {\bibinfo {title} {{Realization of a Kondo Insulator in a Multilayer Moire Superlattice}},\ }\href {https://doi.org/10.48550/arXiv.2507.01329} {\bibfield  {journal} {\bibinfo  {journal} {arXiv e-prints}\ ,\ \bibinfo
  {eid} {arXiv:2507.01329}} (\bibinfo {year} {2025})},\ \Eprint {https://arxiv.org/abs/2507.01329} {arXiv:2507.01329 [cond-mat.str-el]} \BibitemShut {NoStop}%
\bibitem [{\citenamefont {Zhang}\ \emph {et~al.}(2021{\natexlab{b}})\citenamefont {Zhang}, \citenamefont {Devakul},\ and\ \citenamefont {Fu}}]{QAH_DFT_zhang_2021}%
  \BibitemOpen
  \bibfield  {author} {\bibinfo {author} {\bibfnamefont {Y.}~\bibnamefont {Zhang}}, \bibinfo {author} {\bibfnamefont {T.}~\bibnamefont {Devakul}},\ and\ \bibinfo {author} {\bibfnamefont {L.}~\bibnamefont {Fu}},\ }\bibfield  {title} {\bibinfo {title} {Spin-textured chern bands in ab-stacked transition metal dichalcogenide bilayers},\ }\href {https://doi.org/10.1073/pnas.2112673118} {\bibfield  {journal} {\bibinfo  {journal} {Proceedings of the National Academy of Sciences}\ }\textbf {\bibinfo {volume} {118}},\ \bibinfo {pages} {e2112673118} (\bibinfo {year} {2021}{\natexlab{b}})},\ \Eprint {https://arxiv.org/abs/https://www.pnas.org/doi/pdf/10.1073/pnas.2112673118} {https://www.pnas.org/doi/pdf/10.1073/pnas.2112673118} \BibitemShut {NoStop}%
\bibitem [{\citenamefont {Devakul}\ and\ \citenamefont {Fu}(2022)}]{QAH_devakul_2022}%
  \BibitemOpen
  \bibfield  {author} {\bibinfo {author} {\bibfnamefont {T.}~\bibnamefont {Devakul}}\ and\ \bibinfo {author} {\bibfnamefont {L.}~\bibnamefont {Fu}},\ }\bibfield  {title} {\bibinfo {title} {Quantum anomalous hall effect from inverted charge transfer gap},\ }\href {https://doi.org/10.1103/PhysRevX.12.021031} {\bibfield  {journal} {\bibinfo  {journal} {Phys. Rev. X}\ }\textbf {\bibinfo {volume} {12}},\ \bibinfo {pages} {021031} (\bibinfo {year} {2022})}\BibitemShut {NoStop}%
\bibitem [{\citenamefont {Rademaker}(2022)}]{Rademaker_hetero_2022}%
  \BibitemOpen
  \bibfield  {author} {\bibinfo {author} {\bibfnamefont {L.}~\bibnamefont {Rademaker}},\ }\bibfield  {title} {\bibinfo {title} {Spin-orbit coupling in transition metal dichalcogenide heterobilayer flat bands},\ }\href {https://doi.org/10.1103/PhysRevB.105.195428} {\bibfield  {journal} {\bibinfo  {journal} {Phys. Rev. B}\ }\textbf {\bibinfo {volume} {105}},\ \bibinfo {pages} {195428} (\bibinfo {year} {2022})}\BibitemShut {NoStop}%
\bibitem [{\citenamefont {Zhou}\ and\ \citenamefont {Wang}(2022)}]{Chern_FP_PDW}%
  \BibitemOpen
  \bibfield  {author} {\bibinfo {author} {\bibfnamefont {S.}~\bibnamefont {Zhou}}\ and\ \bibinfo {author} {\bibfnamefont {Z.}~\bibnamefont {Wang}},\ }\bibfield  {title} {\bibinfo {title} {Chern fermi pocket, topological pair density wave, and charge-4e and charge-6e superconductivity in kagomé superconductors},\ }\bibfield  {journal} {\bibinfo  {journal} {Nature Communications}\ }\textbf {\bibinfo {volume} {13}},\ \href {https://doi.org/10.1038/s41467-022-34832-2} {10.1038/s41467-022-34832-2} (\bibinfo {year} {2022})\BibitemShut {NoStop}%
\end{thebibliography}%

\clearpage
\begin{widetext}

%%%%%%%%%%
%\appendix 
\setcounter{page}{1} 
\setcounter{equation}{0} 
\setcounter{figure}{0} 
\renewcommand{\figurename}{Supplemental Figure}
\renewcommand{\theequation}{\thesection.\arabic{equation}}

\renewcommand{\thefigure}{S\arabic{figure}}      % printed label e.g. S1, S2, ...
\renewcommand{\theHfigure}{S\arabic{figure}}     % unique hyperlink anchor
%%%%%%%%%%%%%%%%

\begin{center}
    {\bf Supplementary material for ``SU(4) Kondo Lattice in Semiconductor Moir\'e Materials"}\\
    Sunghoon Kim
\end{center}

\section{Schrieffer-Wolff transformation}
%\label{appdx:SW}
In this section, we provide details of the Schrieffer-Wolff transformation. We begin by rewriting Eq.~\ref{eq:SU4Anderson} as
\beq 
\tilde{H}&=& T_{cc} + T + V \nonumber\\ 
T_{cc}&=& -t_c \sum_{i,j,\alpha} (c^\dagger_{i,\alpha}c_{j,\alpha}+\tn{h.c.}) ,\nonumber \\
T &=& -t_f  \sum_{ij,\alpha} \left(f^{\dagger}_{i,\alpha}f_{j,\alpha}+ \tn{h.c.}\right)-t_{\perp}  \sum_{ij,\alpha} \left(c^{\dagger}_{i,\alpha}f_{j,\alpha}+ \tn{h.c.}\right),\nonumber\\
V&=& \frac{U}{2} \sum_i  n^f_i (n^f_i -1 ) + \frac{\Delta}{2} (N_c - N_f) \equiv V_U + V_\Delta .
\eeq 
The kinetic terms involving $f$-fermions can be decomposed as
\beq 
T=\sum_{q=-1}^{1} \sum_{m=-3}^{3} T_{q,m},
\eeq 
where $T_{q,m}$ changes the occupation imbalance $(N_f -N_c)/2$ by $q$ and raises the interaction energy by $mU$. For instance, $T_{1,3}$ corresponds to the hopping $(n_i^c,n_i^f):(1,3)\rightarrow (0,4)$. The $T_{q,m}$ operators satisfy $[V_U,T_{q,m}]=mU T_{q,m}$ and $[V_\Delta,T_{q,m}]=-q\Delta T_{q,m}$.

Our goal is to derive an effective Hamiltonian within the low energy subspace. This can be achieved via the unitary transformation 
\beq 
H_{eff}=e^{iS}\tilde{H}e^{-iS} = \tilde{H}+[iS,\tilde{H}]+\frac{1}{2!}[iS,[iS,\tilde{H}]]+\cdots ,
\eeq 
which eliminates hopping terms that change the interaction energy or the occupation imbalance. To organize the expansion systematically, we define
\beq 
\tilde{H}&\equiv& H'^{(1)},\nonumber\\
H'^{(k+1)}&=& e^{iS^{(k)}} H'^{(k)} e^{-iS^{(k)}},
\eeq 
where $H'^{(k)}$ collects terms up to the $k$th-order of kinetic terms (e.g. $t_f^k/U^{k-1}$). To describe $k$-th order processes, we introduce 
\beq 
T^{(k)}_{\vec{q},\vec{m}}\equiv T_{q_1,m_1}\cdots T_{q_k,m_k} ,
\eeq 
with $\vec{q}=(q_1,\cdots,q_k)$ and $\vec{m}=(m_1,\cdots,m_k)$. These operators satisfy $[V_U,T^{(k)}_{\vec{q},\vec{m}}]=M_{\vec{m}}UT^{(k)}_{\vec{q},\vec{m}}$ and $[V_\Delta,T^{(k)}_{\vec{q},\vec{m}}]=-Q_{\vec{q}}\Delta T^{(k)}_{\vec{q},\vec{m}}$, where $M_{\vec{m}}\equiv\sum_{i=1}^k m_i$ and $Q_{\vec{q}}\equiv\sum_{i=1}^k q_i$. 

Suppose that $H'^{(k)}$ contains  $k$-th order terms with $Q_{\vec{q}}\ne 0$ or $M_{\vec{m}}\ne 0$. Such terms must be eliminated by projection onto the low energy subspace. Denoting these terms as $\sum_{\vec{q},\vec{m}}'C_{\vec{q},\vec{m}}T^{(k)}_{\vec{q},\vec{m}}$, we obtain the recursive relation
\beq 
iS^{(k)}-iS^{(k-1)}=\sum_{\vec{q},\vec{m}}'\frac{C_{\vec{q},\vec{m}}}{M_{\vec{m}}U-Q_{\vec{q}}\Delta}T^{(k)}_{\vec{q},\vec{m}}.
\eeq  
In the low energy effective Hamiltonian, $T^{(k)}_{\vec{q},\vec{m}}$ terms must not lower the interaction energy below its initial value, i.e. $\sum_{i=n}^k m_i \geq 0$ for all $0 \leq n \leq k$. Furthermore, as we focus on the subspace $n_f=1$, the allowed initial and final $(q,m)$ combinations are constrained to $(q_k,m_k)\in \{(1, 1), (0, 1), (-1, 0)\}$, $(q_1,m_1)\in \{(1, 0), (0, -1), (-1, -1)\}$. At high orders, allowed terms can be identified numerically. Up to third order, we obtain 
\beq 
H'^{(3)}&=&T_{cc} + \frac{T_{-1,-1}T_{1,1}}{\Delta-U}-\frac{T_{1,0}T_{-1,0}}{\Delta}-\frac{T_{0,-1}T_{0,1}}{U}+\frac{T_{0,-1}T_{0,0}T_{0,1}}{U^2} + H_{add},
\eeq 
where $H_{add}$ consists of terms not allowed in our model due to the hopping range and lattice geometry:
\beq 
H_{add}&=& \frac{T_{-1,-1}T_{cc}T_{1,1}}{(\Delta-U)^2}+\frac{T_{-1,-1}T_{0,0}T_{1,1}}{(\Delta-U)^2}-\frac{T_{-1,-1}T_{1,0}T_{0,1}}{U(\Delta-U)} \nonumber\\
&-&\frac{T_{0,-1}T_{-1,0}T_{1,1}}{U(\Delta-U)}+\frac{T_{0,-1}T_{1,1}T_{-1,0}}{\Delta U} + \frac{T_{1,0}T_{-1,-1}T_{0,1}}{\Delta U} \nonumber \\
&+&\frac{T_{1,0}T_{cc}T_{-1,0}}{\Delta^2} + \frac{T_{1,0}T_{0,0}T_{-1,0}}{\Delta^2}. 
\eeq 
For instance, the second term in $H_{add}$ describes a process involving three sites, $c_i\rightarrow f_j \rightarrow f_k \rightarrow c_i$, which is not allowed in our model as each $c$-site connects to only one $f$-site. Such terms are allowed and may play an important role on other lattices, such as the honeycomb lattice realized in MoTe$_2$/WSe$_2$.

Dropping $H_{add}$, we obtain the effective Kondo Hamiltonian in terms of fermionic operators
\beq 
\bar{H}&=&T_{cc}+\sum_{i,\alpha,\beta} t_{\perp}^2 \left( \frac{c^\dagger_{i,\beta}f_{i,\beta}f^\dagger_{i,\alpha}c_{i,\alpha}}{\Delta-U} - \frac{f^\dagger_{i,\beta}c_{i,\beta}c^\dagger_{i,\alpha}f_{i,\alpha}}{\Delta}  \right) -\frac{2t^2_f}{U}\sum_{\langle ij \rangle,\alpha,\beta} f^\dagger_{i,\beta}f_{j,\beta}f^\dagger_{j,\alpha}f_{i,\alpha} -\frac{6t_f^3}{U^2}\sum_{\substack{\langle ijk \rangle \in \triangle/\triangledown \\ \alpha,\beta,\gamma}}(f^\dagger_{i,\gamma}f_{k,\gamma}f^\dagger_{k,\beta}f_{j,\beta}f^\dagger_{j,\alpha}f_{i,\alpha} + \tn{h.c.}).
\eeq 

\section{Plaquette order in the $J_3/J_2\rightarrow 0$ limit}
%\label{appdx:plaquette}

\renewcommand{\thefigure}{S1}
\begin{figure}[h]
\centering
\includegraphics[width=0.6\linewidth]{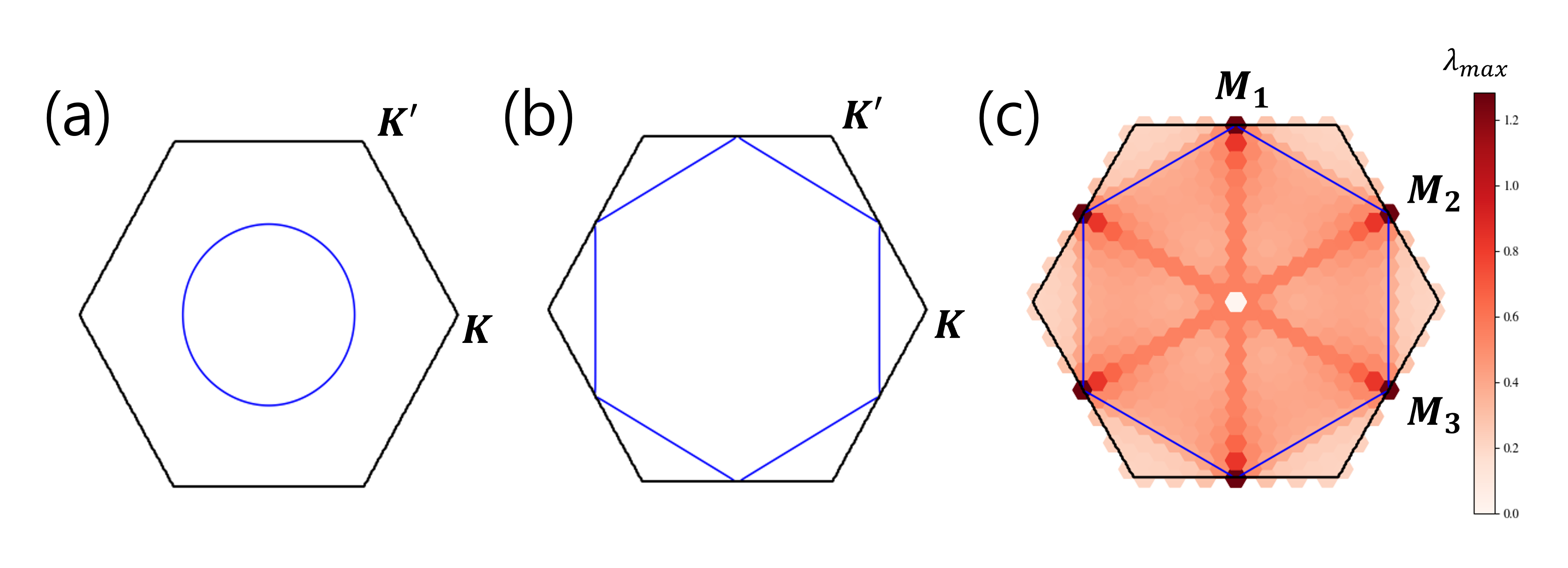} %0.9
\caption{Fermi surface (blue) of the triangular lattice tight-binding model at quarter filling, for (a) electron doping and (b) hole doping. (c) Largest eigenvalue of the polarization tensor across the first Brillouin zone, indicating an instability toward a bond order at the three M points.}
\label{fig:FS_quarter}
\end{figure}

In this section, we argue that the plaquette order at small $|J_3/J_2|$ arises from a nesting-driven instability of triangular-lattice tight-binding model at quarter hole filling. In the $|J_3/J_2|\ll1$ limit, the $f$-fermion Hamiltonian reduces to (flavor indices suppressed) 
\beq 
H_f &=& -\sum_{\langle ij \rangle} \left(\chi_{ij} f^\dagger_i f_j + \tn{h.c.}\right) \nonumber \\
&=& -J_2 \sum_{\vr}\sum_{\tau=1,2,3} \left(A_{\vr,\vr+\ve_\tau} f^\dagger_{\vr+\ve_\tau} f_{\vr} + \tn{h.c.}\right),
\eeq 
where $\ve_{1}=(1,0),\:\ve_{2}=\left(-\frac{1}{2},\frac{\sqrt{3}}{2}\right),\ve_{3}=\left(-\frac{1}{2},-\frac{\sqrt{3}}{2}\right)$ are three lattice vectors (lattice constant set to unity), and $A_{\vr,\vr+\ve_\tau}= \langle f^\dagger_{\vr}f_{\vr+\ve_\tau}\rangle$ is a variational parameter. We initialize a homogeneous (real) ansatz $A_{\vr,\vr+\ve_\tau}=A_0$, where the sign of $A_0$ at the saddle point is \textit{a priori} unknown. At quarter filling, the fermiology is dramatically different between $A_0>0$ (electron doping) and $A_0<0$ (hole doping): the Fermi surface is circular for $A_0>0$, while it is perfectly nested for $A_0<0$ (see Fig.~\ref{fig:FS_quarter}). In the following, we demonstrate that $A_0<0$ is energetically preferred, as nesting drives an instability leading to a gap opening. Introducing a periodic perturbation to the ansatz, $\delta A_\tau (\vQ) e^{i\vQ\cdot\left(\vr+\frac{\ve_\tau}{2}\right)}$, we obtain a perturbation to the Hamiltonian
\beq 
\delta H_f&=&-J_2 \sum_{\vr,\tau}\left( \delta A_\tau(\vQ) e^{i\vQ\cdot\left(\vr+\frac{\ve_\tau}{2}\right)} f^\dagger_{\vr+\ve_\tau} f_{\vr}+\tn{h.c.} \right) \nonumber \\
&=& -J_2 \sum_{\tau} \left( \delta A_{\tau}(\vQ) \sum_{\vk} f^\dagger_{\vk+\vQ}f_{\vk} e^{-i\left(\vk+\frac{\vQ}{2}\right)\cdot \ve_\tau} + \tn{h.c.} \right) \nonumber \\
&\equiv& -J_2 \sum_{\vk} \left(V_{\vk}(\vQ) f^\dagger_{\vk+\vQ}f_{\vk} + \tn{h.c.}\right).
\eeq 
To first order in $V_{\vk}(\vQ)$, the expectation value $\langle f^\dagger_{\vk+\vQ}f_{\vk} \rangle$ varies as
\beq 
\delta \langle  f^\dagger_{\vk+\vQ}f_{\vk} \rangle &=& -J_2 \frac{n_F(E_{\vk})-n_F(E_{\vk+\vQ})}{E_{\vk}-E_{\vk+\vQ}}V_{\vk}(\vQ) \nonumber \\
&=&  -J_2 \frac{n_F(E_{\vk})-n_F(E_{\vk+\vQ})}{E_{\vk}-E_{\vk+\vQ}} \sum_\tau \delta A_{\tau}(\vQ)e^{-i \left( \vk+\frac{\vQ}{2}  \right)\cdot \ve_{\tau} },
\eeq 
where $n_F(E_{\vk})$ is the Fermi-Dirac distribution of the unperturbed dispersion $E_{\vk}$. Multiplying by $\exp\big[i (\vk+\tfrac{\vQ}{2}) \cdot \ve_{\tau'}\big]$ and summing over $\vk$, we have
\beq 
\delta A_{\tau'}(\vQ)=-J_2 \sum_{\vk} \frac{n_F(E_{\vk})-n_F(E_{\vk+\vQ})}{E_{\vk}-E_{\vk+\vQ}} \sum_{\tau} \delta A_{\tau}(\vQ) e^{i \left( \vk+\frac{\vQ}{2} \right)\cdot \ve_{\tau'}} e^{-i \left( \vk+\frac{\vQ}{2} \right)\cdot \ve_{\tau}},
\eeq 
where we have used $\delta A_\tau (\vQ) = \sum_{\vk}\langle f^\dagger_{\vk}f_{\vk+\vQ} \rangle e^{i \left( \vk+\frac{\vQ}{2} \right)\cdot \ve_{\tau}}$. Defining a $3\times 3$ polarization tensor
\beq 
\Pi_{\tau',\tau}(\vQ)\equiv \sum_{\vk} \frac{n_F(E_{\vk+\vQ})-n_F(E_{\vk})}{E_{\vk}-E_{\vk+\vQ}} e^{i \left( \vk+\frac{\vQ}{2} \right)\cdot \ve_{\tau'}} e^{-i \left( \vk+\frac{\vQ}{2} \right)\cdot \ve_{\tau}},
\eeq 
we obtain a self-consistent equation
\beq 
\delta A_{\tau} (\vQ) = J_2 \sum_{\tau'} \Pi_{\tau,\tau'}(\vQ) \delta A_{\tau'}(\vQ).
\eeq 
An instability toward a momentum-$\vQ$ bond order occurs when the largest eigenvalue $\lambda_{max}$ of $\Pi(\vQ)$  satisfies $\lambda_{{max}}\geq 1$. We find that this is indeed the case, as shown in Fig.~\ref{fig:FS_quarter}(c).

\section{CSL and DC for $J_3/J_2<0$}
%\label{appdx:Pijk}

In this section, we demonstrate that the three-site exchange term gives rise to CSL and DC for $J_3<0$. To address the general situation, we apply an external magnetic field to the system. The three-site term can be written as
\beq 
H_3 &=& -J_3 \sum_{\substack{\langle ijk \rangle \in \triangle/\triangledown \\ \alpha,\beta,\gamma}} (f^\dagger_{i,\gamma}f_{k,\gamma}f^\dagger_{k,\beta}f_{j,\beta}f^\dagger_{j,\alpha}f_{i,\alpha} e^{i\Phi_{\tn{ext}}} + \tn{h.c.}), \nonumber \\
&=& -J_3 \sum_{\substack{\langle ijk \rangle \in \triangle/\triangledown \\ \alpha,\beta,\gamma}}  \left( P_{ijk}e^{i\Phi_{\tn{ext}}}+ \tn{h.c.} \right),
\eeq 
where $\Phi_{\tn{ext}}$ is the magnetic flux per triangle. $P_{ijk}=\sum_{\alpha,\beta,\gamma}| \gamma_i \alpha_j \beta_k \rangle\langle\alpha_i \beta_j \gamma_k|$ denotes a three-site ring exchange operator. Note the identity $P_{ijk}=P_{ij}P_{jk}$. Consider the fundamental representation of the SU(N) group, with the generators $\{T^a \}\:(a=1,\cdots,N^2-1)$ being $N\times N$ traceless hermitian matrices. The two-site exchange can be written as $P_{ij} = \frac{1}{N} + 2 T^a_i T^a_j$. The generators satisfy the following commutation and anticommutation relations:
\begin{align}
[T^a, T^b] &= i f_{abc} T^c , \\
\{ T^a, T^b \} &= \frac{1}{N} \delta_{ab} + d_{abc} T^c ,
\end{align}
where $f_{abc}$ and $d_{abc}$ are antisymmetric and symmetric tensors, respectively. Using these relations, the ring exchange can be rewritten as
\beq 
P_{ijk}&=& \frac{1}{N^2}+\frac{2}{N}\left( T^a_i T^a_j + T^a_j T^a_k + T^a_k T^a_i \right) - 2if_{abc}T^a_i T^b_j T^c_k + 2d_{abc}T^a_i T^b_j T^c_k ,
\eeq 
leading to
\begin{align}
P_{ijk} e^{i\Phi_{\tn{ext}}} + \tn{h.c.} 
&= 2\cos{\Phi_{\tn{ext}}} \Bigg\{ \frac{2}{N}\Big( 
      T^a_i T^a_j + T^a_j T^a_k + T^a_k T^a_i \Big) 
      + 2d_{abc} T^a_i T^b_j T^c_k + \frac{1}{N^2} \Bigg\} + 4\sin{\Phi_{\tn{ext}}}f_{abc} T^a_i T^b_j T^c_k .
\end{align}
Let us turn off the external magnetic field $(\Phi_{\tn{ext}}=0)$, as in our model. For SU(2) spins, the ring exchange simply reduces to the sum of two-site couplings since $d_{abc}=0$. On the other hand, $d_{abc}$ are nonzero for $N>2$, and thus the three-site coupling term survives in the SU(4) case. We note that the three-site coupling term is time-reversal even, yet can lead to the CSL that spontaneously breaks time-reversal symmetry, as discussed in main text.

To understand the competition between the Kondo-unscreened phases of the SU(4) Kondo lattice model, consider 
\beq 
H_3
&=&  - J_3 \sum_{\langle ijk \rangle \in \triangle/\triangledown} (T^a_i T^a_j +T^a_j T^a_k +T^a_k T^a_i ) - 4J_3 \sum_{\langle ijk \rangle \in \triangle/\triangledown} d_{abc}T^a_i T^b_j T^c_k + \tn{(const)} \nonumber \\
&=& -2J_3\sum_{\langle ij \rangle} T^a_i T^a_j - 4 J_3 \sum_{\langle ijk \rangle} d_{abc} T^a_i T^b_j T^c_k + \tn{(const)}.
\eeq 
Note that $J_2=1$ and $J_3 = 6t^3_f/U^2<0$ in our model. For sufficiently negative $J_3$, the first term leads to strong bond ordering. Therefore, the decoupled chain phase can be favorable.   

\end{widetext}

\end{document}